\documentclass[conference]{IEEEtran}
\IEEEoverridecommandlockouts

\usepackage{cite}
\usepackage{amsmath,amssymb,amsfonts}
\usepackage{algorithmic}
\usepackage{graphicx}
\usepackage{textcomp}
\usepackage[dvipsnames]{xcolor}
\usepackage{soul}
\usepackage[hyphens]{url}

\usepackage[bookmarks=true,breaklinks=true,letterpaper=true,colorlinks,citecolor=blue,linkcolor=blue,urlcolor=blue]{hyperref}
\usepackage{tcolorbox}
\usepackage{enumitem}

\def\BibTeX{{\rm B\kern-.05em{\sc i\kern-.025em b}\kern-.08em
    T\kern-.1667em\lower.7ex\hbox{E}\kern-.125emX}}

\usepackage{tikz}

\usepackage{nameref}
\newcounter{mylabelcounter}
\makeatletter
\newcommand{\labeltext}[2]{%
#1\refstepcounter{mylabelcounter}%
\immediate\write\@auxout{%
  \string\newlabel{#2}{{here}{\thepage}{{\unexpanded{#1}}}{mylabelcounter.\number\value{mylabelcounter}}{}}%
}%
}
\makeatother

\usepackage{mathtools}

\usepackage{listings}
\definecolor{codegray}{rgb}{0.9, 0.9, 0.9}
\definecolor{codegreen}{rgb}{0, 0.6, 0}
\definecolor{codeblack}{rgb}{0, 0, 0}
\lstdefinestyle{mystyle}{
    language=bash,
    backgroundcolor=\color{codegray},
    commentstyle=\color{codegreen},
    frame=single, 
    framerule=0.75pt,
    rulecolor=\color{codeblack},
    columns=fullflexible,
    basicstyle={\linespread{1}\rmfamily\small},
    lineskip=0.5ex, 
    xleftmargin=.15cm, xrightmargin=.12cm, 
    aboveskip=.15cm, belowskip=.24cm, 
    framextopmargin=.075cm, framexbottommargin=.05cm,
    breaklines, breakautoindent=false, breakindent=2ex, 
}
\usepackage[ruled, linesnumbered, vlined]{algorithm2e}
\usepackage{setspace}
\SetAlFnt{\footnotesize}
\SetAlCapNameFnt{\small}
\SetAlCapFnt{\small}
\IncMargin{3pt}
\SetAlgoNlRelativeSize{-0.5}
\SetKwProg{Def}{def}{$\,$:}{}
\SetKwProg{Func}{Func}{$\,$:}{}
\SetKwInOut{Input}{Input}
\SetKwInOut{Output}{Output}

\usepackage{etoolbox} 
\makeatletter
\patchcmd{\algocf@makecaption@ruled}{\hsize}{\textwidth}{}{} 
\patchcmd{\@algocf@start}{-1.5em}{0em}{}{} 
\makeatother

\usepackage{booktabs, makecell, multirow, tabularx, threeparttable} 
\usepackage{caption}
\usepackage{colortbl}
\definecolor{light-gray}{gray}{0.88}

\newcommand{\workname}{\text{P\textsuperscript{3}-LLM}}
\title{\huge
P\textsuperscript{3}-LLM: An Integrated NPU-PIM Accelerator for \\Edge LLM Inference Using Hybrid Numerical Formats
}
\author{
  \IEEEauthorblockN{
    Yuzong Chen\IEEEauthorrefmark{1},
    Chao Fang\IEEEauthorrefmark{2}, 
    Xilai Dai\IEEEauthorrefmark{1},
    Yuheng Wu\IEEEauthorrefmark{3}, \\
    Thierry Tambe\IEEEauthorrefmark{3},
    Marian Verhelst\IEEEauthorrefmark{2}, 
    Mohamed S. Abdelfattah\IEEEauthorrefmark{1}
  }
  \vspace{4pt}
  \IEEEauthorblockA{
    \IEEEauthorrefmark{1}\textit{Department of Electrical and Computer Engineering, Cornell University}\\
    \IEEEauthorrefmark{2}\textit{EAST-MICAS, KU Leuven}\\
    \IEEEauthorrefmark{3}\textit{Department of Electrical Engineering, Stanford University}\\
  }
  \vspace{5pt}
  \IEEEauthorblockA{
    \IEEEauthorrefmark{1}\{yc2367, mohamed\}@cornell.edu \\
    \IEEEauthorrefmark{2}\{chao.fang, marian.verhelst\}@esat.kuleuven.be \\
    \IEEEauthorrefmark{3}\{yuhengwu, ttambe\}@stanford.edu
  }
  \vspace{-17pt}
}

\begin{document}


\maketitle

\pagestyle{plain}     
\thispagestyle{plain} 


\newcommand{\hpcaheight}{0mm}
\ifdefined\eaopen
\renewcommand{\hpcaheight}{12mm}
\fi

\begin{abstract}
The substantial memory bandwidth and computational demands of large language models (LLMs) present critical challenges for efficient inference. To tackle this, the literature has explored heterogeneous systems that combine neural processing units (NPUs) with DRAM-based processing-in-memory (PIM) for LLM acceleration. However, the high-precision PIM compute units incur significant area and power overhead in DRAM technology, limiting the effective computation throughput. In this paper, we introduce \workname{}, a novel NPU-PIM integrated accelerator for edge LLM inference. Our approach is threefold: First, we propose a flexible mixed-precision quantization scheme, which leverages hybrid numerical formats to quantize different LLM operands with high compression efficiency and minimal accuracy loss. 
Second, we architect an efficient PIM accelerator for \workname{}, featuring enhanced compute units to support hybrid numerical formats. 
Our careful choice of numerical formats allows to co-design low-precision PIM compute units that significantly boost the computation throughput under iso-area constraints. 
Third, we optimize the low-precision dataflow of different LLM modules by applying operator fusion to minimize the overhead of runtime dequantization. 
Evaluations on diverse LLMs and tasks demonstrate that \workname{} achieves higher accuracy than state-of-the-art KV-cache quantization and weight-activation quantization algorithms. Combining the proposed quantization scheme with low-precision PIM architecture co-design, \workname{} yields an average of $4.9\times$, $2.0\times$, and $3.4\times$ speedups over state-of-the-art LLM accelerators HBM-PIM, Ecco, and Pimba, respectively. Code is available at \url{https://github.com/yc2367/P3-LLM}.

\end{abstract}

\section{Introduction} \label{sec:intro}

Large language models (LLMs) have revolutionized various machine learning tasks such as text generation~\cite{llama, mistral, qwen-2p5}, image understanding~\cite{llama-3.2-vision, qwen-vl}, and logical reasoning~\cite{deepseek-r1, openai-o3, llama-4}. 
\textcolor{black}{
\labeltext{Besides}{reb:motivation_edge} large-scale cloud serving, LLMs are becoming increasingly prevalent in diverse edge scenarios such as chatbot interaction, autonomous driving, and mobile assistant~\cite{awq, h2-llm, llm4drive, llama-3p2, mobilellm}. 
However, the intelligence of LLMs comes at the cost of substantial computation and memory demands, imposing a significant bottleneck for low-cost deployment, particularly in edge scenarios with limited hardware resources.
}

\textcolor{black}{To enhance the performance and efficiency of edge LLM deployment, heterogeneous LLM acceleration based on neural processing unit (NPU) and DRAM-based processing-in-memory (PIM) has been actively explored~\cite{neupim, ianus, attacc, samsung-pim, pimba, sk-pim, orches}.}
An integrated NPU-PIM accelerator leverages NPU and PIM to accelerate two distinct LLM inference stages, prefilling and decoding, respectively. During the prefilling stage, the LLM performs compute-intensive general matrix-matrix multiplication (GEMM) that can be effectively accelerated by NPU with high computational parallelism~\cite{tpuv4i}.
While during the decoding stage, the LLM performs memory-intensive general matrix-vector multiplication (GEMV) that is well-suited for PIM with much higher bandwidth than conventional DRAM~\cite{samsung-aim, sk-aim}. 
Despite PIM's great promise in accelerating LLM inference, the PIM compute units (PCUs) designed for high-precision (e.g. \texttt{FP16}) arithmetic suffer from significant area and power overhead due to the much lower transistor density in DRAM technology~\cite{upmem}.
This PCU overhead severely limits the computation throughput of PIM, restricting its practical speedup mainly to single-batch inference on early-generation LLMs, whose multi-head attention and feed-forward layers exhibit an arithmetic intensity of one during decoding~\cite{llama, llama-2}. Recently, low-batch edge LLM inference has gained much popularity~\cite{h2-llm}, and state-of-the-art (SoTA) LLMs have adopted grouped-query attention (GQA) with arithmetic intensities larger than one~\cite{llama-3p1, llama-3p2, qwen-2p5, mistral}. 
This trend necessitates new PCU design with higher throughput, while remaining within DRAM area constraints.

One effective solution to alleviate the cost of compute units is quantization. 
By reducing the operand bit-width, quantization not only decreases the memory footprint, but also enables faster and more area-efficient computation on low-precision hardware. Depending on the target quantized operands, LLM quantization algorithms can be broadly classified into three categories: (1) \textit{weight-only} quantization~\cite{awq, gptq, microscopiq}; 
(2) \textit{KV-cache-only} quantization~\cite{oaken, kivi, kvquant}; (3) \textit{weight-activation}\footnote{Here, we mainly refer activation quantization as quantizing both input activations and KV-cache.} quantization~\cite{tender, quarot, qserve, ecco, anda, mx_plus, mant}. Among them, weight-only and KV-cache-only quantization are known to achieve near-lossless accuracy at 4-bit precision. 
However, since all other operands remain in \texttt{FP16}, these two methods offer limited compression and still require expensive hardware units for computation~\cite{figna, bitmod, figlut}. In contrast, weight-activation quantization significantly reduces both memory footprint and hardware computation cost, yet applying it to enable cost-effective PIM for LLM inference still poses several challenges. 

On the algorithm side, SoTA weight-activation quantization algorithms like QuaRot~\cite{quarot} and QoQ~\cite{qserve} primarily use the conventional integer format, and rely on calibration datasets to approximate the dynamic behavior of activations in a static offline manner, thereby struggling to maintain good accuracy for other datasets due to overfitting. 
On the hardware side, adopting low-precision PCUs for LLM inference remains an open research problem in the architecture community, mainly because of the challenges in balancing memory saving, model accuracy, and hardware efficiency. MANT~\cite{mant} and Ecco~\cite{ecco} propose aggressive quantization methods and co-designed NPU accelerators, but suffer from low computational efficiency due to the complicated numerical formats and encoding to represent quantized operands. Thus, these proposals cannot be directly applied to PIM with stringent area constraints.
While a recent PIM work, Pimba~\cite{pimba}, adopts the 8-bit microscaling format to improve the area efficiency of PCUs, it overemphasizes preserving model accuracy with 8-bit KV-cache-only quantization, resulting in low memory savings for edge LLM inference. 
Furthermore, offloading the self-attention module of LLMs to low-precision PCU introduces additional challenges, such as the difficulty of accurately quantizing attention-scores.
  
To overcome the above limitations, we propose \workname{}, an N\underline{P}U-\underline{P}IM integrated accelerator for low-\underline{P}recision edge \underline{LLM} inference. Given the substantial memory and computation demands of LLMs, \workname{} employs mixed-precision quantization with 4-bit weights and KV-cache to achieve high compression, as well as 8-bit activations and attention scores to reduce computational complexity. Our key algorithmic innovation is an operand-dependent quantization scheme that leverages hybrid numerical formats to aggressively quantize different LLM operands with minimal accuracy loss. For KV-cache, we propose a dynamic, input-aware smoothing strategy that mitigates outliers without calibration datasets, thus enabling accurate and efficient 4-bit integer quantization while preventing overfitting. For attention-scores, we introduce an unsigned 8-bit floating-point format with 4-bit exponent and 4-bit mantissa (\texttt{FP8-S0E4M4}) to offer superior numerical fidelity. For activations, we explore the accuracy-efficiency tradeoffs among different 8-bit quantization options, and identity the direct \texttt{FP8-E4M3} cast as the optimal choice. Our quantization algorithm is further equipped with a low-precision PIM co-design, which features high-throughput and area-efficient PCUs to flexibly support hybrid numerical formats. Finally, we architect low-precision dataflow of different LLM modules by applying operator fusion to minimize the overhead of runtime dequantization. 
  
The main contributions of this paper are summarized below:
\begin{enumerate}
    \item We introduce \workname{}, an algorithm-hardware co-design solution with carefully optimized low-precision dataflow to unleash the potential of PIM for edge LLM inference. 
    \item We propose a novel LLM quantization scheme that employs hybrid numerical formats to achieve an excellent trade-off between model accuracy, memory footprint, and computational efficiency. 
    \item We design a high-throughput and area-efficient PCU architecture to accelerate the proposed quantization scheme, while remaining within the PIM area constraints. 
    \item Through comprehensive evaluation, we demonstrate that \workname{} achieves higher accuracy than SoTA LLM quantization algorithms Oaken~\cite{oaken}, QuaRot~\cite{quarot}, and QoQ~\cite{qserve}, while offering an average of $4.9\times$, $2.0\times$, and $3.4\times$ speedups over the SoTA LLM accelerators HBM-PIM~\cite{samsung-pim}, Ecco~\cite{ecco}, and Pimba~\cite{pimba}, respectively.
\end{enumerate}
%

\section{Background} \label{sec:background}

\subsection{Architecture of Large Language Models (LLMs)} \label{sec:llm_arch}
\begin{figure} [t]
    \centering
    \includegraphics[width=1\linewidth]{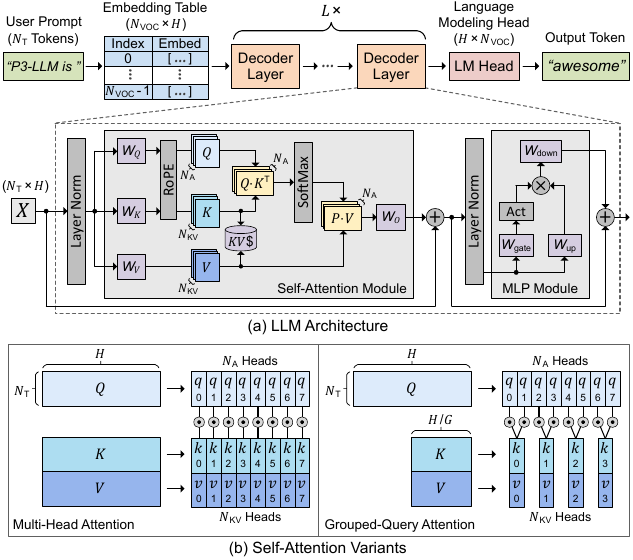}
    \vspace{-14pt}
    \caption{Illustration of LLM architecture.}
    \label{fig:llm_arch}
  \end{figure}

As depicted in Fig.~\ref{fig:llm_arch}(a), mainstream LLMs have a series of decoder layers in addition to an input embedding table and an output language modeling (LM) head. During the prefilling stage of inference, the LLM receives a user prompt containing $N_\text{T}$ tokens, and converts it to an input matrix through an embedding table of size $N_\text{VOC} \times H$, where $N_\text{VOC}$ is the vocabulary size and $H$ is the hidden dimension size. The input matrix is processed by $L\times$ decoder layers, followed by the LM head at the end to produce the first output token. During the decoding stage, the LLM takes this output token as input, and performs the same operation as in the prefilling stage to generate new tokens in an auto-regressive manner. 

The decoder layer serves as the fundamental component in LLMs, consisting of a self-attention module and a multi-layer perceptron (MLP). The self-attention module begins with three linear layers ($W_Q$, $W_K$, $W_V$) to generate query, key, and value vectors, respectively. In recent LLM architectures~\cite{llama, mistral, qwen-2p5}, $W_Q$ and $W_K$ are usually followed by rotary position embedding (RoPE)~\cite{rope}, which encodes positional information into the query and key vectors through matrix rotation. The generated key and value vectors are also cached in off-chip memory for computation reuse during future decoding iterations, and are therefore referred to as the KV-cache. Then, each query and key-value vector is split into $N_\text{A}$ and $N_\text{KV}$ heads, respectively, where $N_\text{A}$ is the number of attention heads and $N_\text{KV}$ is the number of key-value heads. For every attention head, the query vectors are multiplied with the transposed key vectors (${Q \cdot K^\text{T}}$), followed by a softmax function to calculate the attention-scores ($P$). The attention-scores are then multiplied with the value vectors (${P \cdot V}$), and the results are passed through a linear layer ($W_O$) to produce the attention output states. The MLP module contains three linear layers ($W_\text{gate}, W_\text{up}, W_\text{down}$) to produce the MLP output states. 

Considering the KV-cache needs to be stored for every token, its capacity can become significant for long-context scenarios~\cite{longbench}. To mitigate this storage overhead, recent LLMs have adopted the GQA mechanism. As shown in Fig.~\ref{fig:llm_arch}(b), conventional multi-head attention has the number of attention heads equal to the number of key-value heads, i.e., $N_\text{A} = N_\text{KV}$. On the other hand, GQA partitions the $N_\text{A}$ attention heads into $G = N_\text{A}\,/\,N_\text{KV}$ groups (two in this example), and different groups share the same key-value vectors, effectively reducing the KV-cache capacity by $G\times$.

  \begin{figure} [t]
    \centering
    \includegraphics[width=1\linewidth]{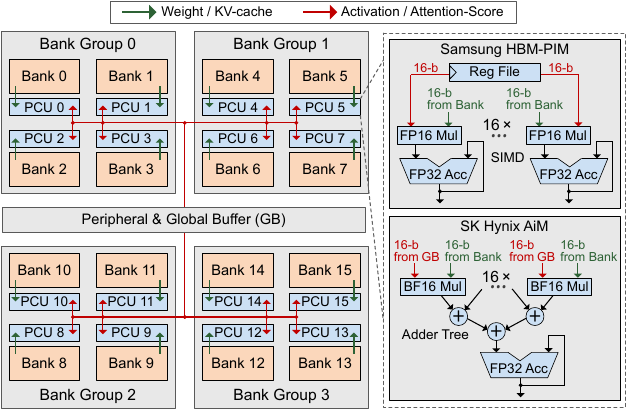}
    \vspace{-12pt}
    \caption{Illustration of PIM architectures for LLM decoding acceleration.}
    \label{fig:pim_arch}
    \vspace{-2pt}
  \end{figure}

\subsection{Processing In-Memory (PIM) for LLM Acceleration}
DRAM-based PIM has become a promising solution to accelerate the decoding stage of LLM inference given its higher internal bandwidth tailored for memory-bound operations~\cite{samsung-pim, sk-pim, attacc, ianus, neupim, pimba, orches}. Fig.~\ref{fig:pim_arch} illustrates the architectures of two commercially available PIM devices: Samsung's HBM-PIM~\cite{samsung-pim} and SK Hynix's Accelerator-in-Memory (AiM)~\cite{sk-pim}. The left part of Fig.~\ref{fig:pim_arch} shows a PIM channel consisting of 16 banks organized into 4 bank groups. One PIM compute unit (PCU) is placed near each DRAM bank to perform efficient GEMV operations by leveraging the abundant bank-level parallelism. Depending on area constraints, two banks may share the same PCU to amortize the area overhead~\cite{samsung-pim}. During LLM decoding, the DRAM bank transfers weights / KV-cache data to the PCU in 256-bit granularity (i.e., 16$\times$16-bit operands). Meanwhile, the input vector is sent from the host to either the PCU register file in HBM-PIM or the global buffer in AiM. 
  
As shown in the right part of Fig.~\ref{fig:pim_arch}, HBM-PIM and AiM have different implementations of the PCU microarchitecture. In HBM-PIM, the PCU contains a 16-way single-instruction-multiple-data (SIMD) MAC unit, and allows to exploit input reuse during GEMV by multiplying the same input element with 16 weights. On the other hand, the PCU of AiM uses the brain floating-point (BF16) format for data representation, and adopts an adder-tree-based design to exploit output reuse during GEMV. Despite their simplicity, the high-precision PCUs incur considerable area overhead, ranging from 20\% to 27\% of the DRAM die area~\cite{samsung-pim, sk-pim}, primarily because the DRAM process has roughly $10\times$ lower transistor density and fewer metal layers for routing compared to CMOS at the same technology node~\cite{upmem}. 
This overhead significantly constrains the achievable compute throughput of PIM, restricting its performance benefits mainly to single-batch inference and multi-head attention that do not exhibit data reuse.

  \begin{figure*} [t]
    \centering
    \includegraphics[width=1\linewidth]{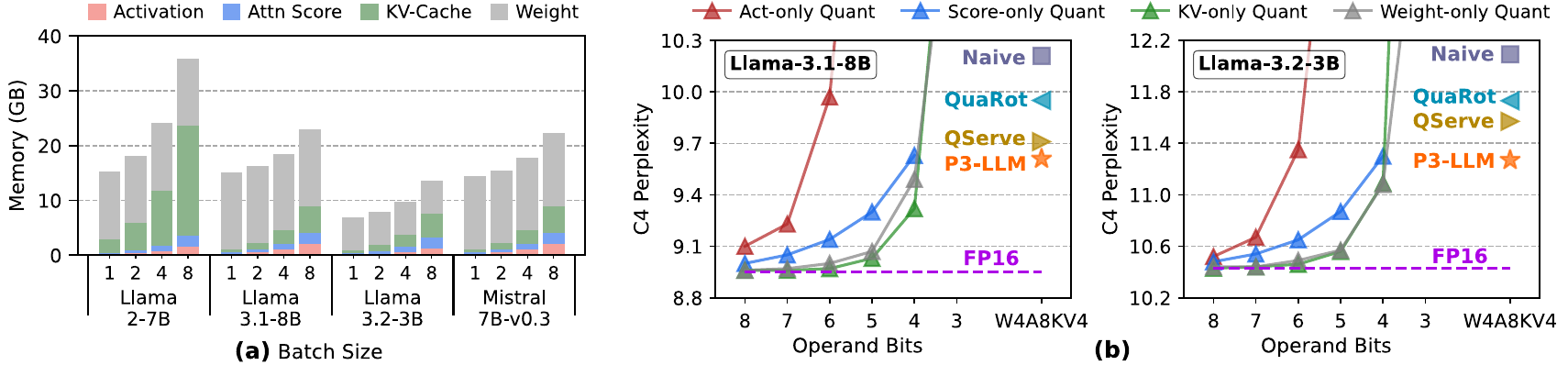}
    \vspace{-12pt}
    \caption{Analysis of LLM operands: (a) Memory footprint of various LLMs at a 4K context length across different batch sizes. (b) Impact of quantization bit-width on the C4 perplexity ($\downarrow$) of Llama-3.1-8B and Llama-3.2-3B. On the x-axis, operand bits represent the precision of quantizing each operand independently. The perplexity of baseline \texttt{FP16} LLMs and different quantization methods under \texttt{W4A8KV4} are also highlighted. Note that all W4A8KV4 methods use \texttt{FP16} attention-scores, except for \workname{}, which uses 8-bit attention-scores.}
    \label{fig:mem_footprint}
    \vspace{-3pt}
  \end{figure*}
  
\subsection{LLM Quantization}
Quantization is a widely used technique for cost-effective LLM acceleration. Consider a group of operands $\mathrm{X}$ and a list of quantization values $\mathrm{Q}$, the quantized operand $\mathrm{X_Q}$ and dequantized operand $\widetilde{\mathrm{X}}$ can be calculated as follows: 
    \begin{equation}
        \Delta = \frac{ |\mathrm{X}|_{\text{max}} }{ \mathrm{Q}_{\text{max}} }  \,; 
        \ \, \mathrm{X_Q} = \texttt{Round}\left(\frac{ \mathrm{X} }{ \Delta }\,,\, \mathrm{Q}\right) ; 
        \ \, \widetilde{\mathrm{X}} = \mathrm{X_Q} \cdot \Delta
        \label{eq:quant_sym}
    \end{equation} 
where $\Delta$ is the scaling factor, and $\texttt{Round}\left(\mathrm{x}, \,\mathrm{Y}\right)$ is a function that rounds a value $\mathrm{x}$ to the closest value in a set $\mathrm{Y}$. This rounding process inevitably introduces error between the original and quantized operands. Numerous techniques have been proposed to reduce quantization error, such as mixed-precision quantization and custom numerical formats. 

In mixed-precision domain, SoTA algorithmic solutions on weight-only~\cite{awq, gptq} and KV-cache-only~\cite{kivi, kvquant} quantization have demonstrated near-lossless accuracy at 4-bit precision. To further alleviate the computation and memory demands of LLMs, several studies have explored weight-activation quantization~\cite{qserve, quarot, omniquant, students}. 
Meanwhile, recent literature in the architecture community has explored custom numerical formats that can better adapt to the tensor distribution of LLMs~\cite{mant, bitmod, blockdialect, fp8, microscaling, olive, mx_plus, spark, amove, microscopiq, anda}. 
In addition to research efforts, custom quantization formats have been widely adopted by industry. For instance, NVIDIA and AMD support the 8-bit floating-point format with two variants: 4-bit exponent 3-bit mantissa (\texttt{FP8-E4M3}) and 5-bit exponent 2-bit mantissa (\texttt{FP8-E5M2})~\cite{fp8}.

\section{Motivation} \label{sec:motivation}

  \begin{figure} [t]
    \centering
    \includegraphics[width=1\linewidth]{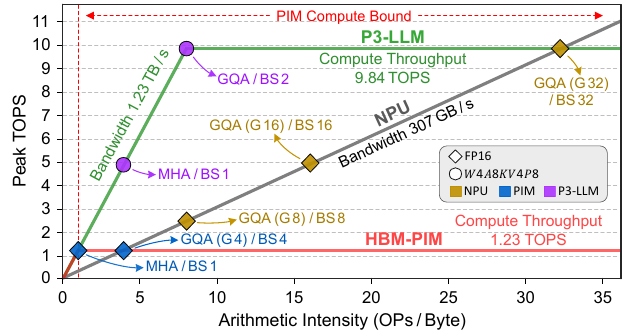}
    \vspace{-13pt}
    \caption{Roofline analysis of different accelerators. The markers highlight the achievable throughput of various operators, including multi-head attention (MHA), grouped-query attention (GQA) with different group sizes (G), and linear layer with different batch sizes (BS).}
    \label{fig:roofline}
  \end{figure}
  
\subsection{Memory Footprint vs. Quantization Sensitivity} \label{sec:mem_vs_quant}
Fig.~\ref{fig:mem_footprint}(a) shows the memory breakdown of various LLMs at \texttt{FP16}, including Llama-2-7B~\cite{llama-2}, Llama-3.1-8B~\cite{llama-3p1}, Llama-3.2-3B~\cite{llama-3p2}, and Mistral-7B~\cite{mistral}. The batch size varies from 1 to 8 and the input context length is 4K, reflecting typical edge LLM inference scenarios~\cite{h2-llm, anda, pimba, zigzag-llm}. While model weights dominate the memory footprint at very low batch sizes, the KV-cache capacity significantly grows with increasing batch sizes. Conversely, activations and attention-scores have a smaller impact on memory footprint, since their memory can be released immediately after the associated GEMM/GEMV modules complete. Furthermore, Llama-2-7B requires much larger KV-cache than other LLMs, due to its usage of multi-head attention. 

Fig.~\ref{fig:mem_footprint}(b) demonstrates the impact of quantizing individual operands to low precision using the standard integer format, while maintaining other operands at \texttt{FP16}. We quantify the model performance of Llama-3.1-8B and Llama-3.2-3B on the C4 dataset~\cite{c4} using the perplexity metric, where lower perplexity indicates better performance. Unlike the trend observed in memory footprint, activations and attention-scores exhibit larger sensitivity to quantization, leading to worse perplexity than weight and KV-cache under the same quantization bit-width. Furthermore, quantizing weight and KV-cache can maintain acceptable perplexity until 4-bit. 

The above observations motivate our mixed-precision quantization scheme, namely \texttt{W4A8KV4P8}. In this scheme, weights (\texttt{W}) and KV-cache (\texttt{KV}) are quantized to 4 bits to maximize memory savings, while activations (\texttt{A}) and attention-scores (\texttt{P}) are retained at 8 bits to mitigate accuracy loss. Although attention-scores account for a small memory footprint, we argue that quantizing attention-scores further enhances hardware efficiency by allowing the $P \cdot V$ operation of the self-attention module to run on low-precision compute units. 
Unfortunately, the na\"ive \texttt{W4A8KV4P8} quantization with the standard integer format results in large perplexity degradation, as highlighted in Fig.~\ref{fig:mem_footprint}(b). To address this, \workname{} proposes an operand-dependent \texttt{W4A8KV4P8} quantization strategy using hybrid numerical formats (detailed in Section~\ref{sec:p3llm_quant}). This approach employs dedicated numerical format for each LLM operand to minimize their quantization error. As a result, \workname{} achieves better model performance than SoTA \texttt{W4A8KV4} integer quantization algorithms, QuaRot~\cite{quarot} and QoQ~\cite{qserve}.

\subsection{Limitations of Existing NPU-PIM Systems} \label{sec:pim_limit}
Existing PIM solutions for LLM inference mainly adopt high-precision PCUs with limited computation throughput, thus facing challenges in low-batch decoding and GQA. To elucidate such limitations, we conduct roofline analysis on HBM-PIM supporting \texttt{FP16} arithmetic~\cite{samsung-pim}, which has $4\times$ higher bandwidth than normal HBM during PIM operations. Fig.~\ref{fig:roofline} reveals that the performance benefits of HBM-PIM gradually disappear as the batch size (BS) and GQA group size (G) approach 4, due to the limited computation throughput. 
On the other hand, NPU remains memory-bound even for moderate BS ${\ge 16}$, making it highly desirable to increase the compute throughput of PIM while exploiting its abundant bandwidth. 
To address this, \workname{} leverages \texttt{W4A8KV4P8} quantization to enable cost-effective LLM inference on compact, low-precision hardware. Compared to HBM-PIM, \workname{} can integrate $4\times$ PCUs under iso-compute-area constraints. Furthermore, as we will discuss in Section~\ref{sec:double_pump}, the low-precision PCU enables $2\times$ higher operating frequency than the \texttt{FP16} PCU, which effectively doubles the peak throughput. Thus, \workname{} offers a superior roofline with $8\times$ higher throughput over HBM-PIM. 
  
  \begin{table} [t!]
    \centering
    \setlength{\tabcolsep}{5.5pt}
    \renewcommand{\arraystretch}{1.2}
    \footnotesize
    \caption{\workname{} vs. existing co-design solutions for quantized LLM acceleration. The precision "W$\alpha$A$\beta$KV$\gamma$P$\delta$" stands for $\alpha$-bit weights, $\beta$-bit activations, $\gamma$-bit KV-cache, and $\delta$-bit attention-scores. By default, attention-scores are 16-bit.}
    \vspace{-4pt}
    \begin{threeparttable}
    \begin{tabular}{ c c c c c }
        \Xhline{0.3ex}
          \multirow{2}{*}{\textbf{Framework}} & \multirow{1}{*}{\textbf{Operand}} & \multirow{1}{*}{\textbf{Memory}} & \multirow{1}{*}{\textbf{Model}} & \textbf{Hardware}  
          \\
          & \textbf{Precision} & \textbf{Saving} & \textbf{Accuracy} & \textbf{Efficiency}  
          \\
        \Xhline{0.3ex}
          BitMoD~\cite{bitmod} & W4A16KV16 & \textcolor{orange}{\textbf{Medium}} & \textcolor{Green}{\textbf{High}} & \textcolor{orange}{\textbf{Medium}} 
          \\ \hline
          Oaken~\cite{oaken} & W16A16KV4 & \textcolor{orange}{\textbf{Medium}} & \textcolor{Green}{\textbf{High}} & \textcolor{Maroon}{\textbf{Low}} 
          \\ \hline
          MANT~\cite{mant} & W4A8KV4 & \textcolor{Green}{\textbf{High}} & \textcolor{orange}{\textbf{Medium}} & \textcolor{orange}{\textbf{Medium}}  
          \\ \hline
          Ecco~\cite{ecco} & W4A8KV4 & \textcolor{Green}{\textbf{High}} & \textcolor{Green}{\textbf{High}} & \textcolor{Maroon}{\textbf{Low}}  
          \\ \hline
          Pimba~\cite{pimba} & W16A16KV8 & \textcolor{Maroon}{\textbf{Low}} & \textcolor{Green}{\textbf{High}} & \textcolor{orange}{\textbf{Medium}} 
          \\ \hline
          \workname{} (Ours) & W4A8KV4P8 & \textcolor{Green}{\textbf{High}} & \textcolor{Green}{\textbf{High}} & \textcolor{Green}{\textbf{High}} 
          \\ \hline
        \Xhline{0.3ex}
    \end{tabular}
    \end{threeparttable}
    \label{tab:overall_comp}
    \vspace{-8pt}
  \end{table}

  \begin{figure*} [t]
    \centering
    \includegraphics[width=1\linewidth]{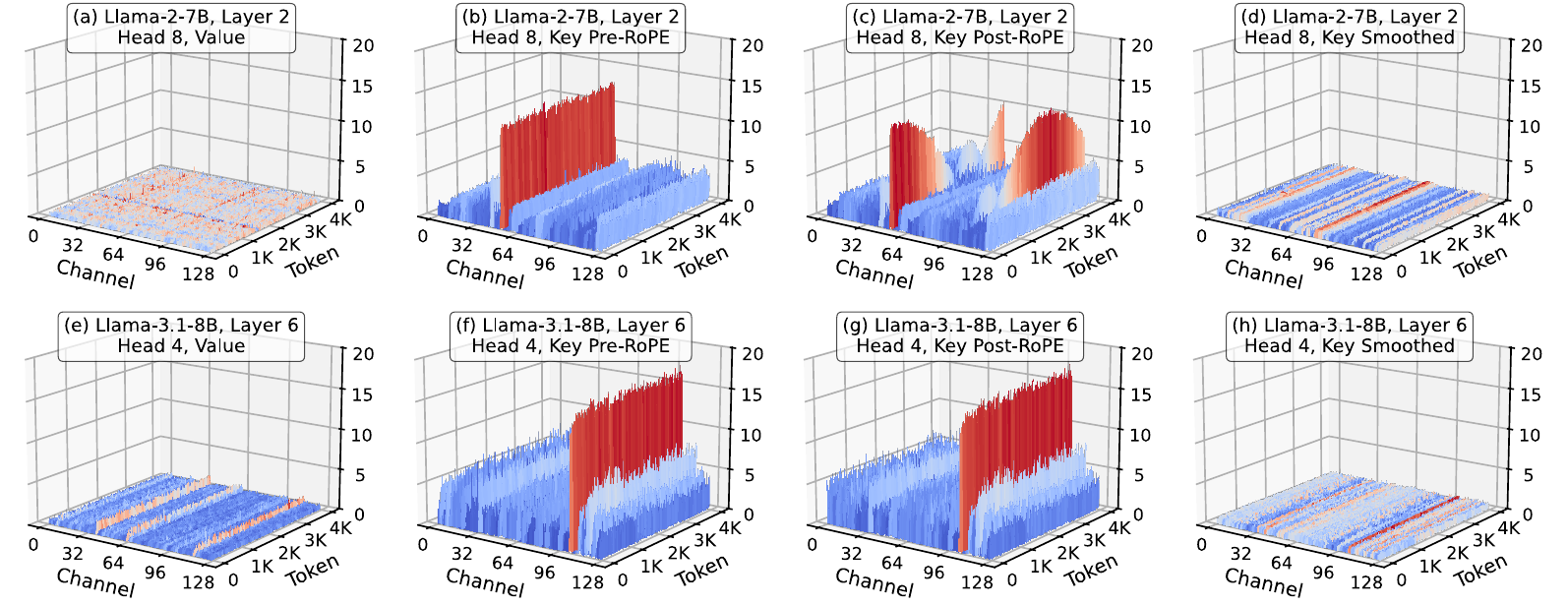}
    \vspace{-12pt}
    \caption{KV-cache distribution (in absolute value) of Wikitext-2 dataset from representative layers and heads of Llama-2-7B and Llama-3.1-8B. The context length is 4K. (a)(e) The value cache shows no outlier pattern. (b)(f) The pre-RoPE key cache shows distinct outlier channels. (c) The post-RoPE key cache of Llama-2-7B contains much less structured outlier pattern. (g) The post-RoPE key cache of Llama-3.1-8B still exhibits distinct outlier channels. (d)(h) \workname{} employs dynamic per-channel smoothing to eliminate the outlier channels of key cache.}
    \label{fig:kv_profile}
    \vspace{-3pt}
  \end{figure*}

\subsection{Limitations of Existing LLM Accelerator Co-Design} \label{sec:codesign_limit}
Numerous co-design solutions have been proposed by the architecture community to accelerate quantized LLMs, as summarized in Table~\ref{tab:overall_comp}.
BitMoD~\cite{bitmod} and Oaken~\cite{oaken} focus on 4-bit quantization of weights and KV-cache, respectively. But since all other operands remain in \texttt{FP16}, these two approaches provide moderate memory saving and hardware efficiency. MANT~\cite{mant} and Ecco~\cite{ecco} achieve greater memory saving and efficiency via \texttt{W4A8KV4} quantization. Nevertheless, MANT exhibits noticeable accuracy drop under this aggressive quantization scheme since it does not address the outlier issue presented in activations and KV-cache. Whereas Ecco requires complicated online decompression of quantized operand back to \texttt{FP16} for computation, leading to considerable hardware overhead. Thus, although Ecco achieves high accuracy, its idea cannot be applied to designing low-precision PCUs for more effieicnt LLM decoding. 
On the other hand, Pimba~\cite{pimba} introduces a low-precision PIM architecture that exploits the 8-bit microscaling format~\cite{microscaling} to mitigate the area overhead of PCUs. However, it leaves the linear layer unquantized, and the \texttt{FP32} accumulation pipeline in the microscaling data path limits the overall hardware efficiency gain. Unlike these proposals, \workname{} seeks to strike a balanced trade-off among memory footprint, model accuracy, and hardware efficiency. By assigning a dedicated precision and efficient numerical format to each LLM operand, \workname{} minimizes both quantization error and memory usage. Our careful choice of hybrid numerical formats allows to co-design low-precision PIM that significantly boosts the computation throughput under iso-compute-area constraints.

\section{\workname{} Quantization Framework} \label{sec:p3llm_quant}

This section details the \workname{} quantization framework, the core of which is a mixed-precision \texttt{W4A8KV4P8} scheme employing hybrid numerical formats. Furthermore, \workname{} introduces a dynamic input-aware smoothing strategy to effectively suppress KV-cache outliers without overfitting.
  

\subsection{KV-Cache Quantization} \label{sec:kv_quant}
\noindent
\textbf{Profiling of KV-cache Distribution.}
Fig.~\ref{fig:kv_profile} illustrates the KV-cache distribution from representative layers and heads of Llama-2-7B and Llama-3.1-8B, using Wikitext-2 dataset~\cite{wikitext} with a context length of 4K.  Fig.~\ref{fig:kv_profile}(a, e) show that the value cache has no outlier pattern, whereas Fig.~\ref{fig:kv_profile}(b, f) reveal that the key cache contains distinct outlier channels that pose significant challenges to quantization. These observations align with previous findings that certain fixed channels in activations exhibit large outliers~\cite{smoothquant, llm-int8}. However, our profiling uncovers an additional important insight: the key cache of Llama-2-7B and Llama-3.1-8B exhibit drastically different patterns after RoPE is applied. In Llama-2-7B, the post-RoPE key cache has much less structured outlier patterns. In contrast, the key cache of Llama-3.1-8B shows little difference before and after RoPE. The reason is that Llama-3 supports a much longer sequence length than Llama-2 (128K vs. 4K tokens). As described in Section~\ref{sec:llm_arch}, RoPE encodes positional information into the key cache via matrix rotation$\,\text{---}\,$specifically, by multiplying each key vector with a rotation matrix whose angle is proportional to the token's position index. When the maximum supported sequence length is long (e.g., 128K in Llama-3), the rotation angles are very small for typical input context lengths (e.g., 4K) targeting edge scenarios~\cite{h2-llm, anda}, and RoPE has minimal impact on the key-cache distribution. Given these insights, we adopt pre-RoPE and post-RoPE key-cache quantization for Llama-2 and Llama-3, respectively, to take the advantage of structured outlier channels.

\noindent
\textbf{\labeltext{Dynamic Key-Cache Smoothing.}{k_smooth_overhead}}
To suppress outlier channels in the key cache ($K$), we propose a dynamic input-aware smoothing strategy. Assume $K$ is of size $N_\text{T} \times H$, where $N_T$ is the input context length and $H$ is the hidden dimension size, the smoothed key cache $K_S$ can be expressed as:
\vspace{-3pt}
    \begin{equation}
        K_S[:,\,c] = \frac{K[:,\,c]}{\texttt{Max}(\,\left|K[:,\,c]\right|\,)} \;{;} \quad \text{for}\; c \in [0, H)
        \label{eq:key_smoothing}
    \end{equation} 
Here, every key channel is divided by a smoothing factor, which is the per-channel absolute maximum, to lie in the numerical range of $[-1, 1]$. Consequently, the outlier channels are effectively suppressed as illustrated in Fig.~\ref{fig:kv_profile}(d, h). 
To apply dynamic smoothing to both prefilling and decoding stages, we calculate the smoothing factors by taking the per-channel absolute maximum of the key cache generated during prefilling. The smoothing factors are then saved and reused in the decoding stage to scale the newly generated key vectors. This approach provides two benefits. First, since every channel stores a smoothing factor shared by all tokens, the additional memory required is inversely proportional to the context length. Given that the context length can easily reach several hundreds, this memory overhead is ${<1\%}$ and negligible. Second, the calculation of smoothing factors only involves the prefilling context and has no impact on the decoding stage. Our profiling on an A6000 GPU running Llama-3.1-8B reveals that this calculation takes ${<5}$ millisecond in total across all layers, even under a context length of 32K. When considering a very stringent service-level objective in chatbot applications, with a 250-millisecond requirement on the time-to-first-token~\cite{distserve}, the runtime overhead of dynamic smoothing is still ${<2\%}$.

While per-channel smoothing has been explored by prior quantization works, our proposed strategy has two notable differences. First, prior methods often rely on a calibration dataset to determine the outlier threshold~\cite{oaken} or smoothing factor~\cite{smoothquant, qserve} offline, which can lead to overfitting. In contrast, our method does not require any calibration dataset. 
\textcolor{black}{Second, unlike prior works that focus exclusively on either pre-RoPE~\cite{kvquant} or post-RoPE~\cite{qserve} key-cache quantization, we investigate both approaches by carefully analyzing the impact of RoPE on the resulting key cache distribution of different models.}
Section~\ref{sec:op_mapping} discuss the impact of pre-RoPE and post-RoPE quantization on the co-designed dataflow of \workname{}.

\vspace{3pt}
\noindent
\textbf{Quantization Format.}
Due to the dynamic nature of KV-cache, we employ 4-bit asymmetric integer (\texttt{INT4-Asym}) quantization, which offers superior computational efficiency. Furthermore, since the value cache shows no outlier pattern and the key cache has been greatly smoothed, their uniform distributions are well-suited to the integer format.

  \begin{table} [t]
      \centering
      \setlength{\tabcolsep}{3pt}
      \renewcommand{\arraystretch}{1.15}
      \footnotesize
      \caption{Wikitext-2 and C4 perplexity ($\downarrow$) using different 8-bit numerical formats to quantize attention-scores. KV-cache is quantized to \texttt{INT4-Asym} following Section~\ref{sec:kv_quant}. The context length is 4K.}
      \vspace{-4pt}
        \begin{tabular}{ c | c  c | c  c | c  c | c  c }
          \Xhline{0.3ex}
            Model & \multicolumn{2}{c|}{Llama-2-7B} & \multicolumn{2}{c|}{Llama-2-13B} & \multicolumn{2}{c|}{Llama-3.1-8B}  & \multicolumn{2}{c}{Llama-3.2-3B} \\
            Dataset & \,Wiki & C4 & \,\,Wiki & C4 & \,\,Wiki & C4 & \,\,Wiki & C4 \\
          \Xhline{0.3ex}
            FP16 & \,5.15 & 6.67 & \,\,4.60 & 6.08 & \,\,5.95 & 8.59 & \,\,7.41 & 10.18 \\
          \hline
            INT8 & \,5.19 & 6.71 & \,\,4.65 & 6.12 & \,\,6.06 & 8.71 & \,\,7.55 & 10.31 
          \\
            FP8-E4M3 & \,5.16 & 6.68 & \,\,4.61 & 6.09 & \,\,5.99 & 8.64 & \,\,7.44 & 10.21 
          \\
            FP8-S0E4M4 & \,\textbf{5.15} & \textbf{6.67} & \,\,\textbf{4.60} & \textbf{6.08} & \,\,\textbf{5.95} & \textbf{8.59} & \,\,\textbf{7.42} & \textbf{10.18}
          \\
          \Xhline{0.3ex}
        \end{tabular}
        \vspace{-10pt}
      \label{tab:p_dtype}
    \end{table}

\subsection{Attention-Score Quantization} \label{sec:p_quant}
\labeltext{In order to fully exploit the hardware efficiency of KV-cache quantization, it is essential to quantize additional operands, e.g., attention-scores, within the self-attention module.}{p_quant_reason} 
Although attention-scores have little impact on the overall memory footprint (see Section~\ref{sec:mem_vs_quant}), they play a critical role in the computation flow. Consider a low-precision PCU supporting 8-bit multiplication, if attention-scores remain in \texttt{FP16}, the PCU cannot be fully utilized to accelerate self-attention. Instead, the quantized value cache must be transferred to NPU and rely on the \texttt{FP16} compute units to perform multiplication with attention-scores, thus diminishing the bandwidth advantages of PIM for LLM decoding.
A straightforward solution is to quantize attention-scores with \texttt{INT8}~\cite{sage-attn}, but this introduces two drawbacks. First, \texttt{INT8} attention-scores cause noticeable perplexity degradation as shown in Table~\ref{tab:p_dtype}. Second, as we will demonstrate in Section~\ref{sec:a_quant}, \texttt{INT8} can lead to significant accuracy loss for activation quantization. Since attention-scores behave the same as activations during the MAC operation on hardware, it is desirable to identify a good numerical format that balances quantization accuracy and hardware complexity. 
    
To address this, we propose an unsigned 8-bit floating-point format, \texttt{FP8-S0E4M4}, that contains a 4-bit mantissa and a 4-bit exponent with an exponent bias of $-15$. There are two insights that drive our design choice. First, because the attention-scores are produced after softmax, its numerical range always lies between $0$ and $1$, eliminating the need for a sign bit in its encoding. Second, recall \texttt{FP16} has a 5-bit exponent with a bias of $-15$, which provides an exponent range of $[-14, 15]$. However, since attention-scores are always less than 1, there is no need to use positive exponent values, leading to an effective exponent range of $[-14, -1]$ with 14 distinct values. Thus, a 4-bit exponent is sufficient to represent the attention-score range, leaving 4 bits to store mantissa with better numerical fidelity. 
Table~\ref{tab:p_dtype} reflects that \texttt{FP8-S0E4M4} outperforms both \texttt{INT8} and \texttt{FP8-E4M3} for attention-score quantization, and achieves near-lossless model performance.

\subsection{Weight Quantization} \label{sec:w_quant}
For weight quantization, we build upon an existing numerical format, BitMoD~\cite{bitmod}, which adaptively remaps the redundant negative zero encoding of \texttt{FP4} to some pre-defined special values. On top of the basic \texttt{FP4} quantization values $\{\pm\,0, \, \pm\,0.5, \, \pm\,1, \, \pm\,1.5, \, \pm2, \, \pm3, \, \pm4, \, \pm6\}$, BitMoD introduces four additional special values $\{\pm\,5, \, \pm\,8\}$, where one of them can be selected to replace the negative zero. By searching for the optimal special value for each group of weights, it reduces the quantization error compared to asymmetric integer quantization with minimal hardware overhead. Notably, since weights and KV-cache are mapped to the same operand on the MAC hardware, a decoder is required to accommodate the numerical formats of both weights and KV-cache (i.e., BitMoD and \texttt{INT4-Asym}), which we discuss in Section~\ref{sec:pcu_design}.

  \begin{table} [t]
    \centering
    \setlength{\tabcolsep}{3pt}
    \renewcommand{\arraystretch}{1.15}
    \footnotesize
    \caption{\textcolor{black}{Wikitext-2 and C4 perplexity ($\downarrow$) of different weight-activation quantization methods. For activations, we compare INT8-based SmoothQuant (SQ) and direct FP8-E4M3 quantization. For weights, we examine 4-bit BitMoD. The context length is 4K.}}
    \vspace{-4pt}
    \begin{tabular}{ c c | c  c | c  c | c  c | c  c }
      \Xhline{0.3ex}
        \multicolumn{2}{c|}{Precision} & \multicolumn{2}{c|}{2-7B} & \multicolumn{2}{c|}{2-13B} & \multicolumn{2}{c|}{3.1-8B}  & \multicolumn{2}{c}{3.2-3B} \\
        Weight & Activation & \,Wiki & C4 & \,\,Wiki & C4 & \,\,Wiki & C4 & \,\,Wiki & C4 \\
      \Xhline{0.3ex}
        16 & 16 & \,5.12 & 6.63 & 4.57 & 6.05 & 5.84 & 8.43 & 7.28 & 10.01 \\
      \hline
        16 & INT8-SQ & 5.15 & 6.67 & 4.61 & 6.09 & 5.92 & 8.54 & 7.34 & 10.08
      \\
        16 & FP8-E4M3 & \textbf{5.12} & \textbf{6.63} & \textbf{4.58} & \textbf{6.05} & \textbf{5.85} & \textbf{8.46} & \textbf{7.31} & \textbf{10.03} \\
      \hline
        4 & INT8-SQ & 5.37 & 6.95 & 4.76 & 6.27 & 6.36 & 9.12 & 7.87 & 10.73 
      \\
        4 & FP8-E4M3 & \textbf{5.24} & \textbf{6.78} & \textbf{4.66} & \textbf{6.15} & \textbf{6.16} & \textbf{8.90} & \textbf{7.64} & \textbf{10.48}
      \\
      \Xhline{0.3ex}
    \end{tabular}
    \label{tab:a_dtype}
    \vspace{-10pt}
  \end{table}

  \begin{figure*}
    \centering
    \includegraphics[width=1\linewidth]{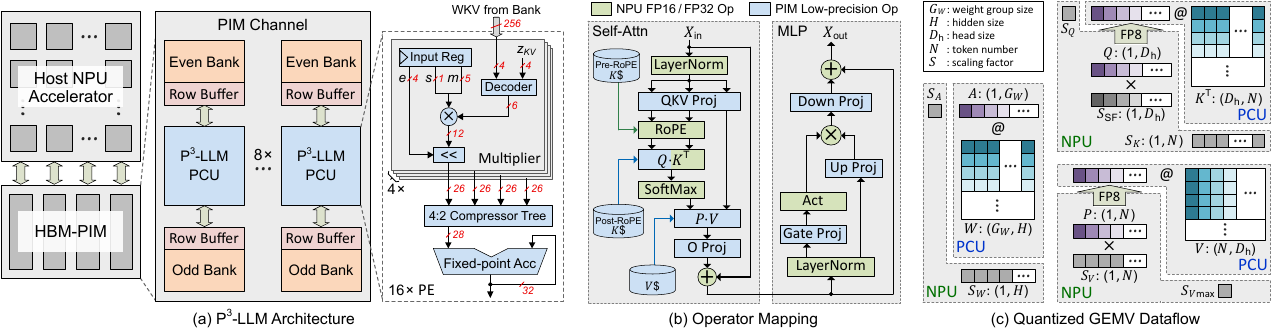}
    \vspace{-12pt}
    \caption{(a) The architecture of \workname{}. (b) Operator mapping on \workname{} during decoding. (c) The quantized dataflow of three GEMV operations: weights $@$ activations, query $@$ key, attention-score $@$ value. For clarity, we use "$@$" and "$\times$" to denote GEMV and element-wise multiplication, respectively. }
    \label{fig:p3_arch}
  \end{figure*}

\subsection{Activation Quantization} \label{sec:a_quant}
Similar to the key cache, activations are also known to exhibit outlier channels~\cite{llm-int8}, which makes standard integer quantization error-prone. 
\textcolor{black}{To reduce the quantization error of integer activations, prior works mainly leverage two techniques to suppress outliers. The first is Hadamard transformation, which applies a Hadamard rotation matrix to activations before performing quantization~\cite{quarot, spinquant}. However, the online Hadamard transformation can bring considerable runtime overhead~\cite{ecco}. The second is smoothing, which migrates the quantization difficulty of outlier activation channels to the corresponding weight channels~\cite{smoothquant}. Unfortunately, the migration of quantization difficulty can significantly increase the error of low-precision weights.}

To address these issues, we propose leveraging an 8-bit format that remains well-suited for activations even in the presence of outliers. Inspired by prior studies~\cite{zeroq_fp, fp8_vs_int8}, we explore \texttt{FP8} as an alternative format for activation quantization without Hadamard transformation and smoothing. We find that the standard \texttt{FP8-E4M3} format can achieve near-lossless accuracy for quantizing activations, owing to its wide numerical range. 
\textcolor{black}{Table~\ref{tab:a_dtype} shows the perplexity of per-token  activation quantization using \texttt{FP8-E4M3} and SmoothQuant-based~\cite{smoothquant} \texttt{INT8}. SmoothQuant causes large perplexity loss under 4-bit weights. This is expected, because weights are already difficult to quantize at 4 bits, yet SmoothQuant migrates additional difficulty to weights. On the other hand, \texttt{FP8-E4M3} significantly outperforms SmoothQuant as its large numerical range can accommodate outliers without affecting weight quantization.}

\section{\workname{} Architecture} \label{sec:p3llm_hardware}
    
This section outlines the architectural details of \workname{}, which features: (a) a high-throughput and area-efficient PCU co-designed to support hybrid numerical formats, (b) an efficient operator mapping that maps memory-intensive operators to the low-precision PIM hardware during decoding, (c) a quantized matrix multiplication flow that applies operator fusion to minimize the runtime overhead of dequantization.

\subsection{\workname{} Microarchitecture} \label{sec:pcu_design}
To support the proposed \texttt{W4A8KV4P8} scheme with hybrid numerical formats in PIM, we design an efficient PCU on top of HBM-PIM~\cite{samsung-pim}. Fig.~\ref{fig:p3_arch}(a) left shows the \workname{} architecture containing a host NPU and an HBM-PIM. One PIM channel contains 8 PCUs, each of them is shared by 2 banks to amortize the area overhead. 
During the decoding stage, the NPU sends activations and attention-scores to the input register of PCU, while weights and KV-cache are read from DRAM banks to perform MAC operations. Given that weights and KV-cache are quantized to 4 bits in \workname{}, we choose to equip each PCU with 64 multipliers to match the 256-bit column access granularity of HBM, effectively quadrupling the compute throughput compared to the baseline HBM-PIM with 16 \texttt{FP16} multipliers. However, scaling up the number of multipliers makes the SIMD-based design of HBM-PIM inefficient, as every multiplier must be paired with a 32-bit accumulator, leading to substantial register overhead. 
    
To address this, we choose to maintain the same number of output registers as in HBM-PIM, and increase the input bandwidth to match the computational throughput. Fig.~\ref{fig:p3_arch}(a) right shows the proposed PCU microarchitecture containing 16 processing elements (PEs). \textcolor{black}{\labeltext{The}{reb:pcu_arch} PCU computes a $1\times4\times16$ GEMV tile\footnote{Through the rest of this paper, we refer a $n \times k \times l$ tile as the dot product between $n \times k$ activations and $k \times l$ weights}, which consumes 4 8-bit inputs, 64 4-bit weights, and produces 16 32-bit outputs. On the other hand, the baseline HBM-PIM's PCU computes a $1\times1\times16$ GEMV tile with one \texttt{FP16} inputs, 16 \texttt{FP16} weights, and 16 \texttt{FP32} outputs. Thus, compared to the baseline HBM-PIM, \workname{} only requires 16-bit additional input register per PCU with minimal overhead.}
Inside the low-precision PCU, every PE computes a 4-way dot product, where the multiplication involves a 6-bit fixed-point multiplier that multiplies the signed ($s$) input mantissa ($m$) with the decoded weight or KV-cache. The multiplier bit-width is determined in the following ways: First, computing the attention-score requires a 5-bit mantissa (including the hidden bit), and the activation requires a sign bit, resulting in a total of 6 bits. Second, since KV-cache adopts \texttt{INT4-Asym} quantization, its precision is 5-bit after adding a 4-bit zero point ($z_\text{KV}$). On the other hand, weights are encoded with the BitMoD data type~\cite{bitmod}, which requires 6 bits in fixed-point representation. After multiplication, the 4-bit input exponent ($e$) is used to shift the product. The results from 4 multipliers are reduced through a ${4:2}$ compressor tree, followed by 32-bit fixed-point accumulation. Thus, the proposed PE eliminates the area-hungry components such as high-precision multipliers and exponent alignment modules in the conventional \texttt{FP16/FP32} MAC unit.

\subsection{PIM Offloading Opportunities with \workname{}} \label{sec:op_mapping}
The proposed \texttt{W4A8KV4P8} quantization allows memory-bound matrix operations to be efficiently executed on the low-precision PCU, as illustrated in Fig.~\ref{fig:p3_arch}(b). During the decoding stage, the host NPU performs element-wise operations such as RoPE and Softmax in high precision, while GEMM/GEMV operators such as linear layers are offloaded to PIM. One exception is ${Q \cdot K^\text{T}}$, whose mapping depends on whether the key cache is quantized pre-RoPE or post-RoPE. In the former case, the quantized key cache lacks positional information during decoding, which can produce incorrect outputs if multiplied directly with the query. Therefore, an online RoPE operation must be applied to the key cache in each decoding iteration. Fortunately, RoPE is essentially a few element-wise operations with negligible runtime overhead\cite{flashinfer}. Since RoPE is performed on NPU and its outputs are high-precision, we opt to calculate the following ${Q \cdot K^\text{T}}$ on NPU if the key cache is quantized before RoPE. In contrast, the post-RoPE quantized key cache can be directly multiplied with the query, which permits ${Q \cdot K^\text{T}}$ to be computed by the low-precision PCU.

\subsection{Optimization on Quantized Computation Flow} \label{sec:quant_dataflow}
\noindent
\textbf{Quantization Granularity.} 
The quantization granularity has a direct impact on the computation flow. For query and KV-cache, we employ per-head quantization, where a group of $D_h$ elements are quantized together and $D_h$ is the head size. This allows immediate quantization of each newly generated token and eliminates the need to buffer a window of \texttt{FP16} KV-cache as done in prior works~\cite{mant, kivi}. Additionally, weights are quantized per-group with a group size of $128$ following BitMoD~\cite{bitmod}. Activations are quantized per-token using \texttt{FP8-E4M3}. Finally, attention-score quantization does not need scaling since \texttt{FP8-S0E4M4} already covers the required numerical range. Instead, we directly round the least significant bits of every \texttt{FP16} attention-score and keep its highest 4 mantissa bits. 

\vspace{3pt}
\noindent
\textbf{\labeltext{Quantized Computation Flow.}{quant_dataflow}}
As shown in Fig.~\ref{fig:p3_arch}(c), \workname{} performs quantized matrix multiplication on the low-precision PCU, and implements operator fusion to minimize the runtime overhead of dequantization, which needs to be executed on NPU using high-precision arithmetic. For linear layers, the dequantization scaling is performed after matrix multiplication. For ${Q \cdot K^{\text{T}}}$, since the post-RoPE key cache contains per-channel smoothing factors ($S_{\text{SF}}$), we fuse $S_{\text{SF}}$ into query via element-wise multiplication prior to \texttt{FP8} quantization. Similarly, for ${P \cdot V}$, we fuse the per-value-head scaling factors ($S_V$) into attention-scores before \texttt{FP8} quantization. One potential issue of fusing $S_V$ is that it may cause the resulting attention-scores' magnitude to become larger than one, which is outside the range of \texttt{FP8-S0E4M4}. To prevent this, we apply a second-level scaling that divides $S_V$ with its maximum $S_{V\text{max}}$, resulting in ${S_V \in [0,1]}$. Note that the quantization scaling factor is always positive by definition in Eq.~\ref{eq:quant_sym}, so fusing $S_V$ still produces unsigned attention-scores. This second-level scaling factor $S_{V\text{max}}$ is later multiplied back into the result of ${P \cdot V}$.

    \begin{figure}
        \centering
        \includegraphics[width=1\linewidth]{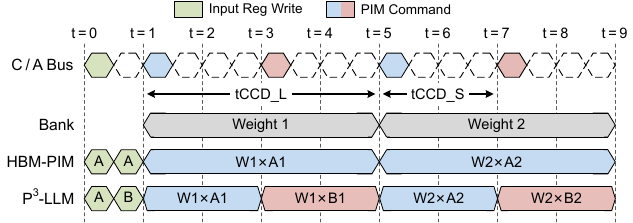}
        \vspace{-14pt}
        \caption{Command timing of HBM-PIM and \workname{}.}
        \label{fig:dp_pcu}
        \vspace{-2pt}
    \end{figure}

\subsection{Throughput-Enhanced PCU} \label{sec:double_pump}
Despite the proposed low-precision PCU achieving $4\times$ the computational throughput than HBM-PIM, it still supports only GEMV operations. To further exploit data reuse opportunities inside the PCU, one can increase the activation tile size from $1\times4$ to $n\times4$ (where $n > 1$), and compute a $n\times4\times16$ GEMM tile. But this will increase the number of multipliers by $n\times$, leading to considerable area overhead. A direct workaround is also to reduce the weight size by $n\times$, resulting in a $n\times4\times(16/n)$ GEMM tile without the need for additional multipliers. Nevertheless, this approach causes severe under-utilization of the PIM bandwidth because of the inherent 256-bit column access granularity of HBM. 

Instead of adopting spatial data, we introduce a novel optimization that exploits temporal data reuse inside the PCU. Our key insight is that the proposed low-precision PCU not only reduces the area, but also permits a higher operating frequency compared to the baseline HBM-PIM~\cite{samsung-pim}. Specifically, our evaluation (detailed in Section~\ref{sec:eval_method}) demonstrates that the PCU of \workname{} can reliably achieve $2\times$ higher frequency than that of HBM-PIM using \texttt{FP32} accumulators. Given that HBM-PIM operates at a clock cycle of $t_{\text{CCD\_L}}$, which corresponds to 4 memory bus cycles, the proposed PCU can instead operate \workname{} at $t_{\text{CCD\_S}}$, which is typically a half of $t_{\text{CCD\_L}}$ across all HBM generations~\cite{hbm-jedec, hbm3-jedec, hbm4-jedec}. 

Fig.~\ref{fig:dp_pcu} compares the command timing between HBM-PIM and \workname{}. Initially, the host NPU sends input activations to the PCU register. Subsequently, a PIM command is initiated every $t_{\text{CCD\_L}}$ on HBM-PIM, which reads out 256-bit weights from the DRAM row buffer and performs a GEMV operation. \workname{} allows the same weight slice to be reused twice by two different inputs, effectively doubling the throughput for cases where multiple inputs share the same weight. 
\textcolor{black}{\labeltext{Our}{reb:discussion_power} simulation shows that the PIM power of \workname{} increases by $28\%$ when its PCU operates at $t_{\text{CCD\_S}}$ instead of $t_{\text{CCD\_L}}$, resulting in $1.56\times$ better energy efficiency. This is because the majority of PIM power dissipation comes from DRAM cell access and column decoder switching, which do not change as the throughput-enhanced PCU reuses the same set of weights from the row buffer across two consecutive $t_{\text{CCD\_S}}$ windows.}

  \begin{table*} [t]
    \centering
    \setlength{\tabcolsep}{4.5pt}
    \renewcommand{\arraystretch}{1.1}
    \small      
    \caption{Wikitext-2 and C4 perplexity ($\downarrow$) under different quantization methods and precisions. For both KV-cache-only and weight-activation quantization, we highlight the best perplexity results in bold.}
    \vspace{-4pt}
    \begin{threeparttable}
    \begin{tabular}{ c c c  c c  c c  c c  c c | c }
        \Xhline{0.3ex}
          \multirow{2}{*}{\textbf{$\ $Dataset$\ $}} & \multirow{2}{*}{\textbf{$\ $Method\tnote{$\dagger$}$\ $}} & \multirow{2}{*}{\textbf{$\ $Precision}} & \multicolumn{2}{c}{\textbf{Llama-1}} & \multicolumn{2}{c}{\textbf{Llama-2}} & \multicolumn{2}{c}{\textbf{Llama-3}} & \multicolumn{2}{c|}{\textbf{Mistral}} & \multirow{2}{*}{\textbf{Mean\,$\Delta$ppl}} 
        \\
          & & & $\;$\textbf{$\ \;$7B$\ \;$} & \textbf{$\ \;$13B$\ \;$}$\;$ & $\;$\textbf{$\ \;$7B$\ \;$} & \textbf{$\ \;$13B$\ \;$}$\;$ & $\;$\textbf{3.1-8B} & \textbf{3.2-3B} & $\;\;$\textbf{7B-v0.1} & \textbf{7B-v0.3} & 
        \\
        \Xhline{0.3ex}
          \multirow{7}{*}{Wikitext} & Baseline & $\ $FP16 & $\;$5.68 & 5.09$\;$ & $\;$5.47 & 4.88$\;$ & $\;$6.24 & 7.81 & $\;\;$5.25 & 5.32 & 0
        \\
        \cline{2-12}
          & Oaken & $\ $KV4 & $\;$5.75 & 5.15$\;$ & $\;$5.53 & 4.93$\;$ & $\;$6.37 & 8.01 & $\;\;$\textbf{5.28} & 5.35 & 0.08
          \\
          & \workname{} & $\ $KV4 & $\;$\textbf{5.72} & \textbf{5.12}$\;$ & $\;$\textbf{5.51} & \textbf{4.92}$\;$ & $\;$\textbf{6.35} & \textbf{7.96} & $\;\;$5.29 & \textbf{5.35} & \textbf{0.06}
          \\
        \cline{2-12}
          & QuaRot & $\ $W4A8KV4 & $\;$5.92 & 5.28$\;$ & $\;$5.70 & 5.07$\;$ & $\;$6.80 & 8.48 & $\;\;$5.41 & 5.49 & 0.30
          \\
          & QoQ & $\ $W4A8KV4 & $\;$5.89 & 5.24$\;$ & $\;$5.69 & 5.07$\;$ & $\;$6.79 & 8.56 & $\;\;$5.42 & 5.50 & 0.30
          \\
          & \workname{} & $\ $W4A8KV4P8 & $\;$\textbf{5.81} & \textbf{5.22}$\;$ & $\;$\textbf{5.65} & \textbf{5.01}$\;$ & $\;$\textbf{6.75} & \textbf{8.39} & $\;\;$\textbf{5.41} & \textbf{5.48} & \textbf{0.25}
          \\ 
        \Xhline{0.3ex}
          \multirow{7}{*}{C4} & Baseline & $\ $FP16 & $\;$7.08 & 6.61$\;$ & $\;$6.97 & 6.47$\;$ & $\;$8.96 & 10.43 & $\;\;$7.74 & 7.83 & 0
          \\
        \cline{2-12}
          & Oaken & $\ $KV4 & $\;$7.15 & 6.66$\;$ & $\;$7.04 & 6.52$\;$ & $\;$9.10 & 10.69 & $\;\;$7.79 & 7.87 & 0.09
          \\
          & \workname{} & $\ $KV4 & $\;$\textbf{7.11} & \textbf{6.63}$\;$ & $\;$\textbf{7.01} & \textbf{6.50}$\;$ & $\;$\textbf{9.09} & \textbf{10.62} & $\;\;$\textbf{7.78} & \textbf{7.87} & \textbf{0.07}
          \\
        \cline{2-12}
          & QuaRot & $\ $W4A8KV4 & $\;$7.31 & 6.77$\;$ & $\;$7.26 & 6.68$\;$ & $\;$9.95 & 11.73 & $\;\;$8.05 & 8.16 & 0.48
          \\
          & QoQ & $\ $W4A8KV4 & $\;$7.28 & 6.75$\;$ & $\;$7.22 & 6.65$\;$ & $\;$9.72 & 11.57 & $\;\;$7.94 & 8.04 & 0.38
          \\
          & \workname{} & $\ $W4A8KV4P8 & $\;$\textbf{7.23} & \textbf{6.72}$\;$ & $\;$\textbf{7.14} & \textbf{6.59}$\;$ & $\;$\textbf{9.64} & \textbf{11.31} & $\;\;$\textbf{7.94} & \textbf{8.03} & \textbf{0.31}
          \\
        \Xhline{0.3ex}
    \end{tabular}

    \begin{tablenotes}
      \item[\textbf{$\dagger$}] For weight-activation quantization, QuaRot and QoQ maintain query and attention-scores in \texttt{FP16}. Whereas \workname{} quantizes attention-scores to \texttt{FP8-S0E4M4} across all models, and quantizes query to \texttt{FP8-E4M3} for Llama-3 and Mistral. 
    \end{tablenotes}
    \end{threeparttable}
    \label{tab:ppl_main}      
    \vspace{-6pt}
  \end{table*} 

\section{Evaluation} \label{sec:evaluation}

\subsection{Experimental Methodology} \label{sec:eval_method}
\noindent
\textbf{Models and Datasets.} 
We select eight representative models from Llama and Mistral families with varying sizes. Specifically, we evaluate \text{Llama-1}-(7B, 13B)~\cite{llama}; \text{Llama-2}-(7B, 13B)~\cite{llama-2}; \text{Llama-3.1}-8B~\cite{llama-3p1}; \text{Llama-3.2}-3B~\cite{llama-3p2}; and \text{Mistral-7B}-(v0.1, v0.3)~\cite{mistral}. We obtain the pre-trained models from their HuggingFace repository, and implement the proposed \texttt{W4A8KV4P8} quantization algorithm in PyTorch. We apply pre-RoPE key-cache quantization to Llama-1 and Llama-2 given their short sequence length, while Llama-3 and Mistral adopt post-RoPE key-cache quantization. For accuracy evaluation of quantized models, we measure the perplexity\footnote{\labeltext{For perplexity comparison with other baseline works}{ppl_footnote}, we use a context length of 2K to align with their reported results. Additionally, Llama-1 supports a maximum context length of 2K; exceeding this limit results in infinite perplexity across all baselines.} on Wikitext-2~\cite{wikitext} and C4~\cite{c4} datasets. Additionally, we evaluate the accuracy on three difficult logical and mathematical reasoning tasks, including MMLU~\cite{mmlu}, ARC-Challenge~\cite{arc-c}, and GSM8K~\cite{gsm8k}. For these reasoning tasks, we use the instruction-tuned variants of Llama-3.1-8B and Llama-3.2-3B as suggested in the LM-Eval framework~\cite{lm-eval}. 

\vspace{3pt}
\noindent
\textbf{Algorithm Baselines.}
We compare \workname{} with SoTA quantization algorithms, including Oaken~\cite{oaken}, QuaRot~\cite{quarot}, and QoQ~\cite{qserve}. Oaken is a 4-bit KV-cache quantization method that determines the key-cache outlier threshold offline using a calibration dataset. We compare Oaken with \workname{} under KV-cache-only quantization for a fair comparison. 
QuaRot and QoQ implement aggressive \texttt{W4A8KV4} quantization using the standard integer format. For all baseline algorithms, we use their official GitHub repository to reproduce the accuracy. 

\vspace{3pt}
\noindent
\textbf{Hardware Implementation.} 
\textcolor{black}{
\labeltext{We}{reb:hw_eval_methodology} model \workname{}'s PCU at RTL-level using SystemVerilog and synthesize it using Synopsys Design Compiler under TSMC 28nm technology. We scale the area and power results from our RTL simulation to HBM-PIM's 20nm technology using DeepScaleTool~\cite{deepscale}.}
We further scale the area overhead by taking into account the effects of DRAM process, which has $10\times$ lower transistor density than the logic process under the same feature size~\cite{upmem}. For performance evaluation, we develop a cycle-level simulator to model the \workname{} system with 4 NPU cores and 16 pseudo HBM channels. The NPU design is based on~\cite{neupim}, where each core contains a 128$\times$128 systolic array, a 128-way vector processing unit, and 16MB on-chip scratchpad that is modeled with CACTI~\cite{cacti}. The PIM subsystem is simulated following the methodology of Newton~\cite{newton}, which is extended to model the computation flow of the proposed PCU. \textcolor{black}{We set the operational frequency of NPU and PCU as 1 GHz and 500 MHz, respectively, considering $t_{\text{CCD\_S}}$ of HBM2 that is 2 DRAM clock cycles.}

\vspace{3pt}
\noindent
\textbf{\labeltext{Accelerator Baselines.}{acc_baseline}}
To evaluate hardware performance, we compare \workname{} against three baselines: 
(1) An NPU accelerator running \texttt{FP16} models, without PIM support in its DRAM; 
(2) An NPU accelerator integrated with \text{HBM-PIM}~\cite{samsung-pim} running \texttt{FP16} models; 
(3) A SoTA LLM accelerator, Ecco~\cite{ecco}, employing \texttt{W4A8KV4}  quantization with k-means codebooks and Huffman encoding. 
All baseline systems share the same NPU and DRAM configurations as \workname{} with comparable total area. 
For comparison of accelerator performance, we focus on the decoding stage that typically dominates the end-to-end inference latency of edge applications~\cite{awq, zigzag-llm}. We use a context length of 4K that covers most edge scenarios as suggested in prior works~\cite{h2-llm, anda}. Notably, for accelerator evaluation, we only assess \text{Llama-2}-(7B, 13B), \text{Llama-3.1}-8B, \text{Llama-3.2}-3B, and \text{Mistral-7B}-v0.3, since the remaining models share the same size and/or architecture.

\subsection{Model Performance} \label{sec:eval_accuracy}
\noindent
\textbf{Perplexity Results.} 
Table~\ref{tab:ppl_main} summarizes the model perplexity of different quantization methods. For KV-cache quantization, on average, \workname{} outperforms Oaken and incurs only ${<0.1}$ perplexity loss. Oaken also has a higher effective KV-cache precision of $4.8$ bits due to its inefficient offline calibration, which keeps a large ratio (${\sim10\%}$) of outliers in high precision. On the other hand, \workname{} employs per-head \texttt{INT4-Asym} quantization, where every 128 elements (i.e., the head dimension size of all evaluated models) share a 16-bit scaling factor and a 4-bit zero-point, resulting in an effective precision of $4.16$ bits. Hence, \workname{} achieves better perplexity even under a lower precision, demonstrating the effectiveness of our dynamic input-aware smoothing for mitigating outliers. For weight-activation quantization, on average, \workname{} outperforms QuaRot and QoQ with $39\%$ and $22\%$ lower perplexity loss, respectively, despite performing more aggressive quantization with 8-bit queries and attention-scores. This highlights the strength of employing our operand-dependent, hybrid numerical formats in preserving model quality under aggressive quantization.

    \begin{table} [t]
      \setlength{\tabcolsep}{3pt}
      \renewcommand{\arraystretch}{1.1}
      \footnotesize
      \vspace{3pt}
      \caption{Accuracy ($\uparrow$) of reasoning tasks under different quantization methods. Oaken and the top \workname{} use \texttt{KV4} quantization. QuaRot and QoQ use \texttt{W4A8KV4} quantization, while the bottom \workname{} uses \texttt{W4A8KV4P8} quantization.}
      \vspace{-4pt}
      \centering
        \begin{tabular}{ c | c c c | c c c | c }
          \Xhline{0.3ex}
            \multirow{1}{*}{Model} & \multicolumn{3}{c|}{Llama-3.1-8B} & \multicolumn{3}{c|}{Llama-3.2-3B} & \multirow{2}{*}{Avg.} \\
            \multirow{1}{*}{Dataset} & MMLU & ARC-C & GSM8K & MMLU & ARC-C & GSM8K \\
          \Xhline{0.3ex}
            FP16 & 72.30 & 84.38 & 84.50 & 63.26 & 76.71 & 76.64 & 76.30
          \\
          \hline
            Oaken & 67.31 & 83.70 & 81.80 & 59.93 & \textbf{76.79} & 75.66 & 74.20
          \\
            \workname{} & \textbf{71.04} & \textbf{83.87} & \textbf{83.02} & \textbf{61.57} & 76.02 & \textbf{75.89} & \textbf{75.24}
          \\
          \hline
            QuaRot & 67.95 & 81.65 & 79.98 & 56.11 & 72.35 & 68.69 & 71.12
          \\
            QoQ & 67.46 & 81.31 & 79.53 & 54.11 & 72.18 & 69.22 & 70.64
          \\
            \workname{} & \textbf{69.22} & \textbf{81.74} & \textbf{82.03} & \textbf{60.05} & \textbf{75.01} & \textbf{74.09} & \textbf{73.69}
          \\
          \Xhline{0.3ex}
        \end{tabular}
        \vspace{-5pt}
      \label{tab:accuracy}
    \end{table}

\vspace{3pt}
\noindent
\textbf{Accuracy Results.}
Table~\ref{tab:accuracy} shows the accuracy of different quantization methods. On average, \workname{} achieves $1.04\%$ higher accuracy than Oaken under 4-bit KV-cache quantization. 
Regarding weight-activation quantization, \workname{} improves the average accuracy by a large margin of $2.57\%$ and $3.05\%$ compared to QuaRot and QoQ, respectively. Moreover, the accuracy improvement is particularly pronounced on Llama-3.2-3B. Both QuaRot and QoQ use integer formats that do not fit well to LLM tensor distributions, and rely on calibration datasets that lead to severe overfitting, particularly for smaller models with limited knowledge capacity.

    \begin{table} [t]
      \setlength{\tabcolsep}{4pt}
      \renewcommand{\arraystretch}{1.125}
      \footnotesize
      \vspace{3pt}
      \caption{Ablation study on the quantization techniques of \workname{} (gray-shaded). The evaluation metric is Wikitext-2 perplexity ($\downarrow$).}
      \vspace{-4pt}
      \centering
        \begin{tabular}{ l | c l | c l }
          \Xhline{0.3ex}
            Quantization Method & \multicolumn{2}{c|}{Llama-2-7B} & \multicolumn{2}{c}{Llama-3.1-8B} 
          \\
          \Xhline{0.3ex}
            Baseline FP16 & 5.47 &  & 6.24 & 
          \\
          \hline
            \ + Pre-RoPE INT4 KV-cache quant & 5.58 & \textcolor{Maroon}{$(\uparrow0.11)$} &    6.52 & \textcolor{Maroon}{$(\uparrow0.28)$}
          \\
            \ + Post-RoPE INT4 KV-cache quant & 5.61 & \textcolor{Maroon}{$(\uparrow0.03)$} &    6.52 & \textcolor{ForestGreen}{$(\cdot)$}
          \\
           \cellcolor{light-gray}{\ \ \ \ $\rightarrow$ Dynamic key-cache smoothing} & \cellcolor{light-gray}{5.51} & \cellcolor{light-gray}{\textcolor{ForestGreen}{$(\downarrow0.10)$}} & \cellcolor{light-gray}{6.35} & \cellcolor{light-gray}{\textcolor{ForestGreen}{$(\downarrow0.17)$}}
          \\
            \ + INT4 weight quant & 5.64 & \textcolor{Maroon}{$(\uparrow0.13)$} &    6.79 & \textcolor{Maroon}{$(\uparrow0.44)$}
          \\
            \cellcolor{light-gray}{\ \ \ \ $\rightarrow$ 4-bit BitMoD weight quant} & \cellcolor{light-gray}{5.63} & \cellcolor{light-gray}{\textcolor{ForestGreen}{$(\downarrow0.01)$}} &    \cellcolor{light-gray}{6.72} & \cellcolor{light-gray}{\textcolor{ForestGreen}{$(\downarrow0.07)$}}
          \\
            \ + FP8-E4M3 attn-score quant & 5.67 & \textcolor{Maroon}{$(\uparrow0.04)$} &    6.81 & \textcolor{Maroon}{$(\uparrow0.09)$}
          \\
            \cellcolor{light-gray}{\ \ \ \ $\rightarrow$ FP8-S0E4M4 attn-score quant} & \cellcolor{light-gray}{5.63} & \cellcolor{light-gray}{\textcolor{ForestGreen}{$(\downarrow0.04)$}} &    \cellcolor{light-gray}{6.73} & \cellcolor{light-gray}{\textcolor{ForestGreen}{$(\downarrow0.08)$}}
          \\
            \ + INT8 activation quant & 5.72 & \textcolor{Maroon}{$(\uparrow0.09)$} & 6.84 & \textcolor{Maroon}{$(\uparrow0.11)$}
          \\
            \cellcolor{light-gray}{\ \ \ \ $\rightarrow$ FP8-E4M3 activation quant} & \cellcolor{light-gray}{5.65} & \cellcolor{light-gray}{\textcolor{ForestGreen}{$(\downarrow0.07)$}} &    \cellcolor{light-gray}{6.75} & \cellcolor{light-gray}{\textcolor{ForestGreen}{$(\downarrow0.09)$}}
          \\
          \Xhline{0.3ex}
        \end{tabular}
        \label{tab:algo_abl}
      \vspace{-10pt}
    \end{table}

\vspace{3pt}
\noindent
\textbf{Algorithm Ablation Study.}
We conduct ablation studies on Llama-2-7B and Llama-3.1-8B to evaluate the accuracy gain of different quantization techniques used in \workname{}. As shown in Table~\ref{tab:algo_abl}, we start with pre-RoPE and post-RoPE KV-cache quantization, and observe noticeable perplexity degradation. This is because both cases do not address the outlier issue of key cache. The proposed dynamic key-cache smoothing significantly improves the perplexity of Llama-2-7B and Llama-3.1-8B by $0.10$ and $0.17$, respectively. Then, adopting BitMoD for 4-bit weight quantization recovers some perplexity loss of \texttt{INT4} weights. Subsequently, quantizing attention-scores to the proposed \texttt{FP8-S0E4M4} format has negligible impact on perplexity. However, applying per-token activation quantization with \texttt{INT8} brings another major perplexity degradation, which is finally recovered by the adopted \texttt{FP8-E4M3} activation quantization. To summarize, \workname{} explores optimal numerical formats that better adapt to different operands, thus minimizing their quantization error. In the presence of key-cache outliers, \workname{} employs a novel smoothing technique to mitigate their impact.

\begin{figure}[t!]
        \centering
        \includegraphics[width=1\linewidth]{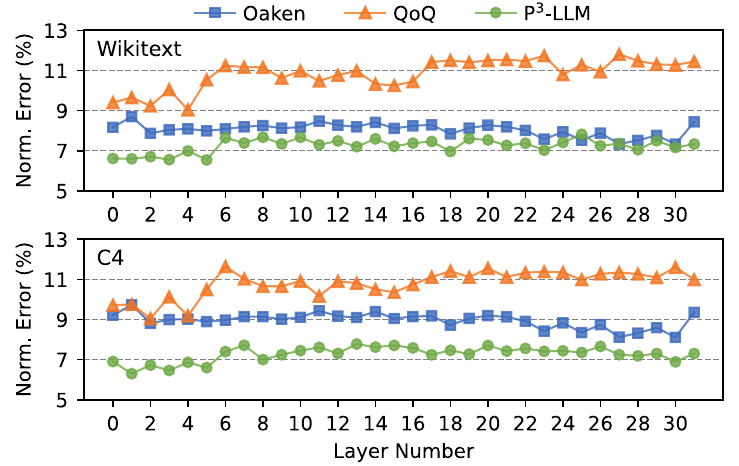}
        \vspace{-15pt}
        \caption{Normalized layer-wise key-cache quantization error of Llama-2-7B on Wikitext-2 and C4 datasets. Oaken and QoQ use Wikitext-2 and Pile as calibration datasets, respectively.}
        \label{fig:k_quant_error}
    \end{figure}

\vspace{3pt}
\noindent
\textbf{\labeltext{Analysis of Key-cache Quantization Error.}{k_quant_error}}
Fig.~\ref{fig:k_quant_error} presents the normalized key-cache quantization error of Oaken, QoQ, and \workname{} on the Wikitext-2 and C4 datasets across all layers of Llama-2-7B. For each layer, we normalize the error to the average magnitude of all samples in the dataset. Oaken and QoQ rely on Wikitext-2 and Pile~\cite{pile} as calibration datasets to address the key-cache outliers. Although Oaken has similar quantization error as \workname{} on Wikitext-2, its error becomes more pronounced on C4 due to overfitting. QoQ has the highest error since the calibration statistics obtained from Pile does not generalize well to both Wikitext-2 and C4. In contrast, the dynamic input-aware smoothing of \workname{} yields the lowest quantization error, as it does not overfit to any calibration dataset.

    \begin{figure*}[t]
        \centering
        \includegraphics[width=1\linewidth]{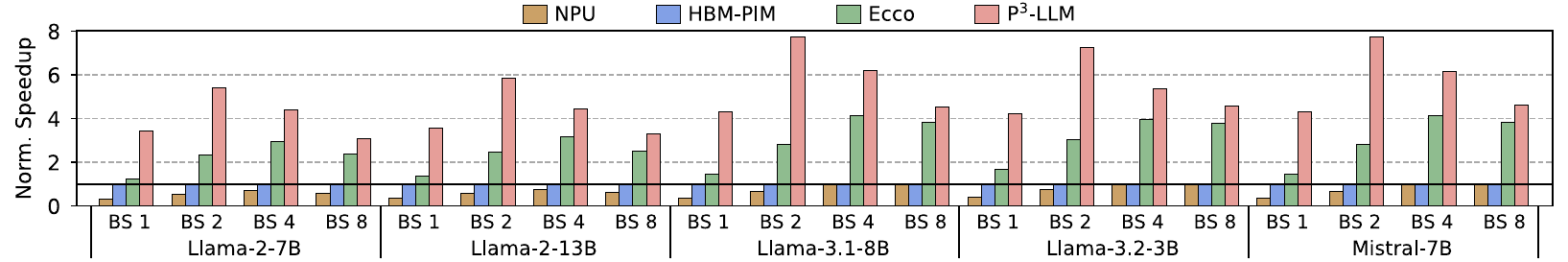}
        \vspace{-15pt}
        \caption{Normalized speedup ($\uparrow$) vs. batch size (BS) for different accelerator systems. The context length is 4K.}
        \label{fig:speedup}
        \vspace{3pt}
    \end{figure*}

    \begin{figure*}[t]
        \centering
        \includegraphics[width=1\linewidth]{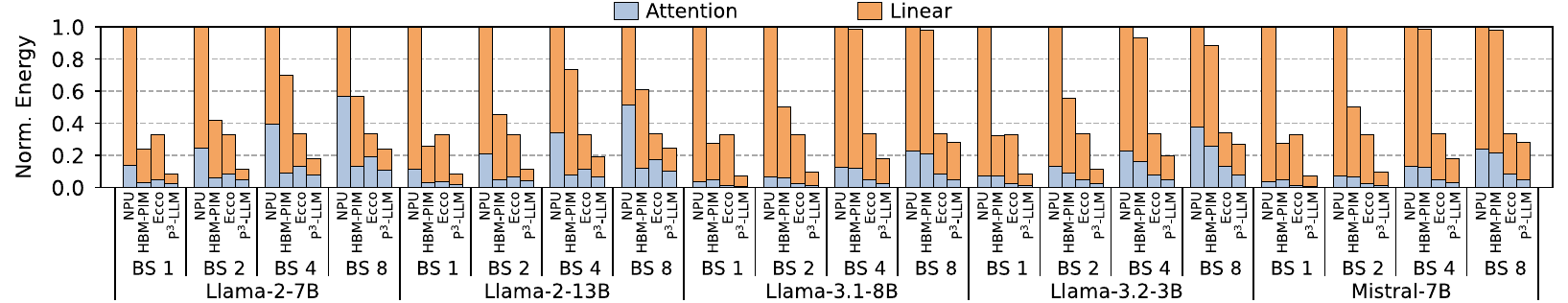}
        \vspace{-15pt}
        \caption{Normalized energy consumption ($\downarrow$) vs. batch size (BS) and the breakdown for attention and linear layers. The context length is 4K.}
        \label{fig:energy}
        \vspace{3pt}
    \end{figure*}

\subsection{Hardware Performance} \label{sec:eval_hardware}
\noindent
\textbf{Speedup.} 
Fig.~\ref{fig:speedup} depicts the normalized decoding speed of \workname{} against three baseline accelerators with batch sizes varying from 1 to 8. The \text{HBM-PIM} system outperforms \text{NPU} at low batch sizes of 1 and 2 across all models, primarily due to its higher internal bandwidth. However, as the batch size reaches 4, the performance advantage of \text{HBM-PIM} gradually diminishes for Llama-2 and even disappears for Llama-3 and Mistral. This is because the linear layer exhibits more data reuse with increasing batch sizes, which \text{NPU} can exploit to boost performance. Additionally, the GQA mechanism of Llama-3 and Mistral offer inherent data reuse opportunities that \text{HBM-PIM} fail to exploit. Furthermore, both \text{NPU} and \text{HBM-PIM} deliver lower performance than Ecco that leverages quantization to reduce the demand of memory bandwidth. 
On the other hand, \workname{} offers substantial performance gains over all baseline accelerators, yielding average speedups of $7.8\times$, $4.9\times$, and $2.0\times$ over \text{NPU}, \text{HBM-PIM}, and Ecco, respectively. These performance gains of \workname{} stem from its careful algorithm-hardware co-design of mixed-precision quantization and efficient PCU architecture. Interestingly, \workname{} demonstrates its highest speedup at a batch size of 2, owing to its throughput-enhanced PCU that allows processing two input vectors within the same $t_{\text{CCD\_L}}$ window.

\vspace{3pt}
\noindent
\textbf{Energy Consumption.}
Fig.~\ref{fig:energy} depicts the breakdown of energy consumption for attention and linear layers across different accelerators and batch sizes. On average, \workname{} yields $6.3\times$, $3.5\times$, and $2.1\times$ better energy efficiency over \text{NPU}, \text{HBM-PIM}, and Ecco, respectively. These energy savings are attributed to the reduced memory footprint offered by the \texttt{W4A8KV4P8} quantization scheme, as well as a PIM architecture co-design that allows most layers to be accelerated by the low-precision PCU. As the batch size increases, \text{NPU} can take advantage of data reuse by loading the model weights only once and processing all input requests within a batch simultaneously. In contrast, the PIM accelerator of \text{HBM-PIM} only supports GEMV, requiring the model weights to be repetitively fetched from the DRAM bank for processing each input request. Thus, the energy consumption of linear layers in \text{HBM-PIM} increases significantly with larger batch sizes.
Compared to \text{HBM-PIM}, the throughput-enhanced PCU in \workname{} enables each memory access to be reused twice, effectively reducing the overhead of duplicated DRAM row activations and the associated energy consumption.

    \begin{figure}[t]
        \centering
        \includegraphics[width=1\linewidth]{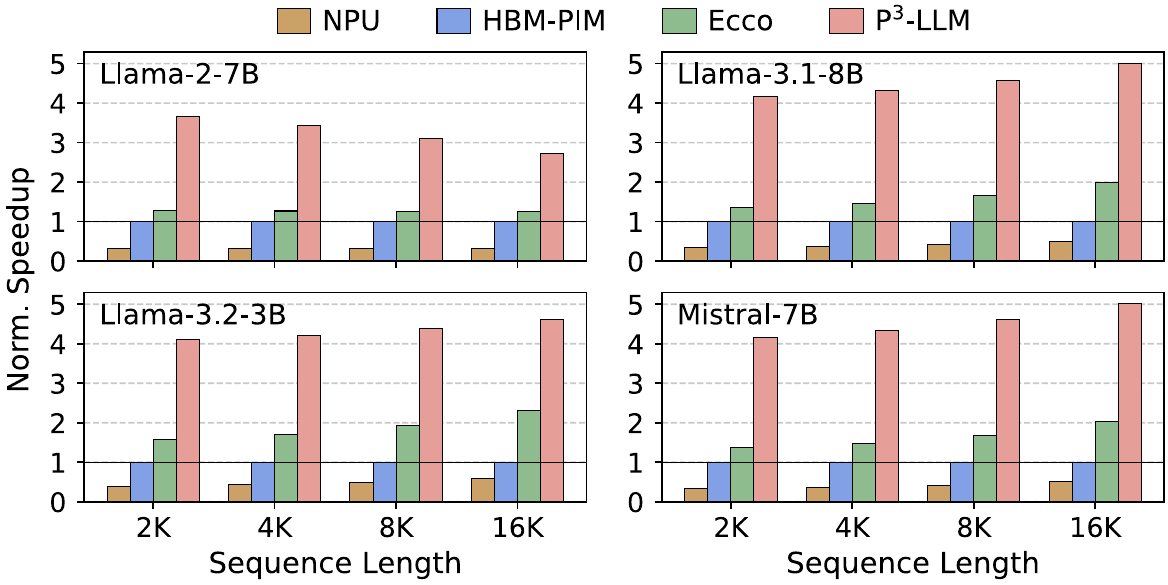}
        \vspace{-15pt}
        \caption{\textcolor{black}{Normalized decoding speedup ($\uparrow$) across different context lengths under single-batch inference.}}
        \label{fig:seqlen_abl}
    \end{figure}

\vspace{3pt}
\noindent
\textbf{\labeltext{Sensitivity to Context Length.}{reb:seq_length}}
\textcolor{black}{Single-batch inference remains one of the most important scenarios for edge LLM deployment. We analyze the impact of context length on the single-batch decoding performance of different accelerator systems. As shown in Fig.~\ref{fig:seqlen_abl}, scaling the context length from 2K to 16K yields additional performance gains for \workname{} across all models, except Llama-2-7B. With a longer context length, the attention layer starts to dominate the overall runtime due to the increased KV-cache footprint. \workname{} effectively mitigates this overhead through its 4-bit KV-cache quantization and attention offloading to the mixed-precision PIM accelerator. For Llama-2-7B, since \workname{} applies key-cache quantization before RoPE, the ${Q \cdot K^{\text{T}}}$ computation is offloaded to NPU, resulting in slightly lower speedup under longer context length.}

\begin{figure}
        \centering
        \includegraphics[width=1\linewidth]{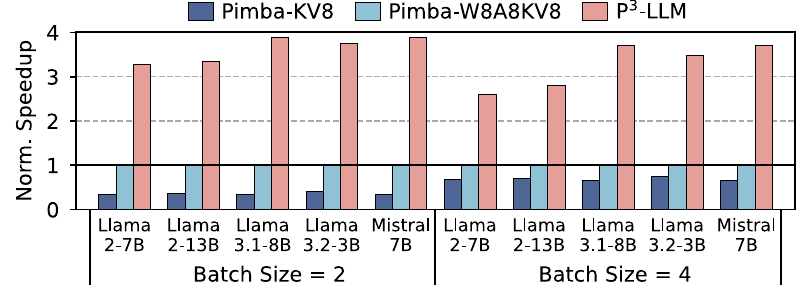}
        \vspace{-15pt}
        \caption{Normalized decoding speedup ($\uparrow$) of Pimba and \workname{}. The context length is 4K.}
        \label{fig:speedup_pimba}
        \vspace{-3pt}
    \end{figure}

\vspace{3pt}
\noindent
\textbf{Comparison with Existing Low-Precision PIM.}
We compare the performance of \workname{} with Pimba~\cite{pimba}, a SoTA low-precision PIM architecture adopting the 8-bit microscaling format~\cite{microscaling}. The original Pimba employs KV-cache-only quantization since it targets cloud serving scenarios, where KV-cache dominates the memory footprint. Hence, we also examine an enhanced version of Pimba with 8-bit weight-activation quantization. Fig.~\ref{fig:speedup_pimba} presents the normalized decoding speed of \workname{} and Pimba under batch sizes of 2 and 4. The original Pimba has the lowest performance because at small batch sizes, weights can dominate the overall memory footprint but remain unquantized. 
By adopting 8-bit weight-activation quantization, on average, the enhanced Pimba achieves $2.1\times$ better performance compared to its original design. The proposed \workname{} further yields an average of $3.4\times$ performance boost compared to the enhanced Pimba. By quantizing weights and KV-cache to 4 bits with minimal accuracy loss, \workname{} reduces the memory access compared to Pimba. Moreover, the throughput-enhanced PCU of \workname{} enables temporal input reuse to double the computational throughput, facilitating efficient execution of low-batch linear layers and GQA.

\begin{figure}[t]
        \centering
        \includegraphics[width=1\linewidth]{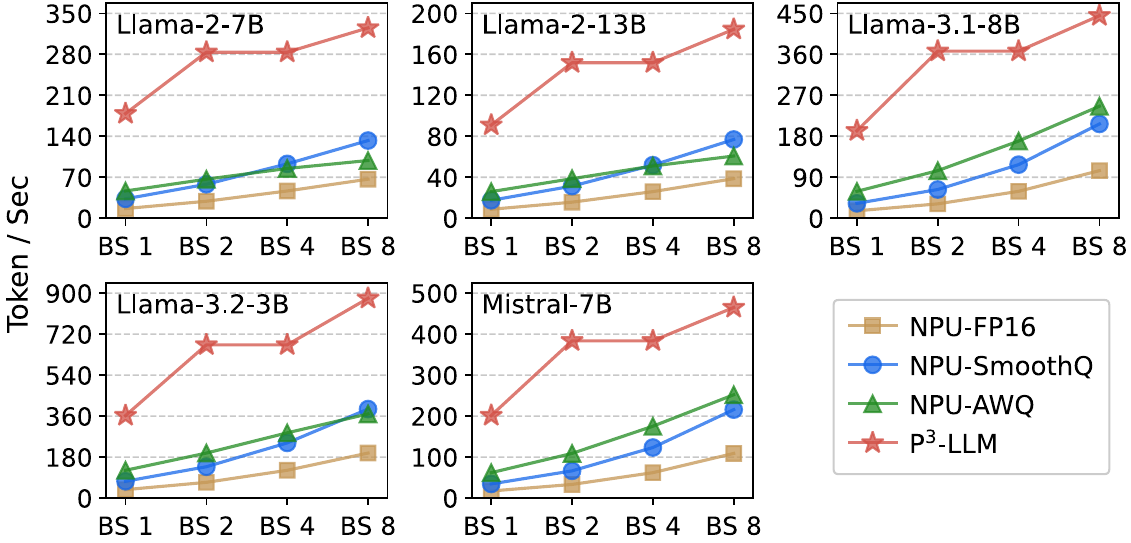}
        \vspace{-12pt}
        \caption{\textcolor{black}{Decoding throughput ($\uparrow$) vs. batch size (BS) of different quantization methods. The context length is 4K.}}
        \label{fig:decoding_throughput}
        \vspace{3pt}
    \end{figure}

\vspace{3pt}
\noindent
\textbf{\labeltext{Comparison with Software Quantization.}{reb:eval_npu_quant}} \textcolor{black}{
We compare \workname{} with two SoTA software quantization algorithms, SmoothQuant~\cite{smoothquant} and AWQ~\cite{awq} running on our baseline NPU. Fig.~\ref{fig:decoding_throughput} illustrates the decoding throughput of different methods across various batch sizes from 1 to 8. On average, \workname{} yields $3.9\times$ and $3.0\times$ higher throughput than SmoothQuant and AWQ, respectively. The performance gains of \workname{} arise from its carefully co-designed \texttt{W4A8KV4P8} quantization algorithm and low-precision PIM architecture. Compared to \workname{}, SmoothQuant and AWQ adopt more conservative 8-bit weight-activation quantization and 4-bit weight-only quantization, respectively. Thus, both methods provide moderate reductions in memory traffic during decoding, which is further exacerbated by the limited off-chip DRAM bandwidth.
}

    \begin{figure}[t]
        \centering
        \includegraphics[width=1\linewidth]{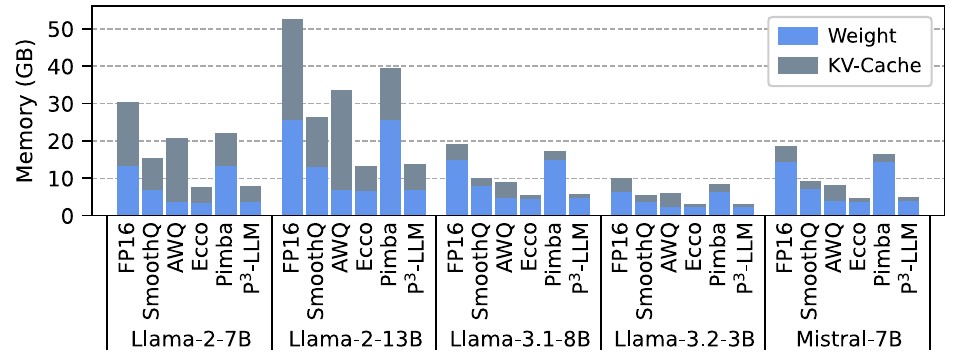}
        \vspace{-15pt}
        \caption{\textcolor{black}{Memory consumption ($\downarrow$) of different quantization methods during decoding. The batch size is 8 and the context length is 4K.}}
        \label{fig:mem_saving}
        \vspace{1pt}
    \end{figure}

\vspace{3pt}
\noindent
\textbf{\labeltext{Memory Analysis.}{reb:mem_saving}} \textcolor{black}{
We examine the memory consumption of weights and KV-cache across various quantization methods during the decoding phase. As shown in Fig.~\ref{fig:mem_saving}, Ecco and \workname{} achieve substantial memory reductions of $3.8\times$ and $3.7\times$ over the \texttt{FP16} baseline, respectively, owing to their aggressive 4-bit quantization of weights and KV-cache that dominate the memory access during LLM decoding. Compared to \workname{}, Ecco has a slightly smaller memory footprint due to its compact Hoffman encoding of quantization codebooks and metadata. However, this advantage is largely negated by the constrained bandwidth between NPU and DRAM, leading to much lower decoding speed than \workname{}.
}

\vspace{3pt}
\noindent
\textbf{\labeltext{Area and Power.}{pcu_area_power}}
Following established methodologies~\cite{pimba, duplex}, we quantify the total area overhead of \workname{} and compare to that of HBM-PIM.
\textcolor{black}{\labeltext{Note}{reb_c:hbm_area_overhead} that the original HBM-PIM described in~\cite{samsung-pim} supports diverse operations such as element-wise multiplication and addition. Given that LLM decoding is dominated by MAC operations, we only model necessary MAC hardware in HBM-PIM, eliminating complicated instruction decoder and unnecessary operand registers.}
As shown in Table~\ref{tab:p3_area_overhead}, \workname{} introduces a total HBM area overhead of $17.5\%$, well below the $25\%$ maximum logic ratio recommended by prior works~\cite{newton, samsung-pim}. Compared to HBM-PIM using \texttt{FP16} arithmetic, \workname{} incurs only $1.1\%$ larger area, yet this increase is justified by delivering an average of $4.9\times$ speedup for edge LLM inference, while maintaining usable accuracy.

We also compare the \workname{} PE with two mixed-precision PEs of SoTA LLM accelerators, BitMoD~\cite{bitmod} and MANT~\cite{mant}. 
Table~\ref{tab:pe_results} shows the area and power of different PE designs under 1GHz frequency. BitMoD exhibits the lowest hardware efficiency, as it requires an expensive \texttt{FP32} accumulator to handle unquantized activations. 
While MANT employs weight-activation quantization, its adaptive numerical type decomposes the weight-activation multiplication into two partial sums with high bit-width. This necessitates an expensive adder to add the two partial sums before accumulation, resulting in large area and energy overhead. \workname{} delivers superior performance over the \texttt{FP16} PE of HBM-PIM, with $3.8\times$ higher energy efficiency per MAC. This substantial efficiency gain is attributed to our efficient quantization approach that reduces the bit-width for both operands and intermediate computation. 

    \begin{table}[t!] 
        \footnotesize
        \renewcommand{\arraystretch}{1.2}
        \setlength{\tabcolsep}{5.5pt}
        \caption{Comparison between HBM-PIM and \workname{}.} 
        \vspace{-3pt}
        \centering 
        \begin{threeparttable}
        \begin{tabular}{ccccc} 
        \Xhline{0.3ex}
          & \multicolumn{2}{c}{\textbf{PCU Area [mm$^2$]}} & \multirow{1}{*}{\textbf{HBM Area}} & \multirow{1}{*}{\textbf{Avg. Norm.}} \\  
          & \textbf{Compute} & \textbf{Buffer} & \textbf{Overhead$\,(\downarrow)$} & \multirow{1}{*}{\textbf{Speedup$\,(\uparrow)$}} \\
        \Xhline{0.3ex}
          HBM-PIM & 7.7 & 6.2 & $16.4\%$ & $1.0\times$ \\ 
        \hline
          \workname{} (Ours) & 8.4 & 6.2 & $17.5\%$ & $4.9\times$ \\ 
        \Xhline{0.3ex}
        \end{tabular}%
        \end{threeparttable}
        \label{tab:p3_area_overhead} 
        \vspace{1pt} 
    \end{table}
    
    \begin{table}[t!] 
        \renewcommand{\arraystretch}{1.15}
        \setlength{\tabcolsep}{4pt}
        \caption{The area and energy consumption of different PE design under 1GHz. The numbers are normalized to that of an FP16 MAC.} 
        \vspace{-3pt}
        \centering 
        \resizebox{0.485\textwidth}{!}{%
        \begin{threeparttable}
        \begin{tabular}{lcll} 
        \Xhline{0.3ex}
        \textbf{Type} & \textbf{MAC$\,$/$\,$Cycle} & \textbf{Area [$\mu$m$^2$]} & \textbf{Energy [pJ$\,$/$\,$MAC]} \\ 
        \Xhline{0.3ex} 
        HBM-PIM & 1 MAC & 1023.1 (1.00$\times$) & 0.69 (1.00$\times$) \\ 
        \hline 
        \multicolumn{1}{l}{MANT} & 2 MACs\tnote{$\,\ddagger$} & 717.3$\,\;$ (0.70$\times$) & 0.40 (0.58$\times$) \\ 
        \multicolumn{1}{l}{BitMoD} & 2 MACs\tnote{$\,\ddagger$} & 1291.6 (1.26$\times$) & 0.61 (0.88$\times$) \\ 
        \multicolumn{1}{l}{\workname{} (Ours)} & 4 MACs\tnote{$\,\ddagger$} & 1109.2 (1.08$\times$) & 0.18 (0.26$\times$) \\ 
        \Xhline{0.3ex}
        \end{tabular}%

        \begin{tablenotes}
          \item[\textbf{$\ddagger$}] Normalized to MACs$\,$/$\,$Cycle under 4-bit weight quantization. 
        \end{tablenotes}
        \end{threeparttable}
        } 
        \label{tab:pe_results} 
    \end{table}


    \begin{figure}[t!] 
        \centering
        \includegraphics[width=1\linewidth]{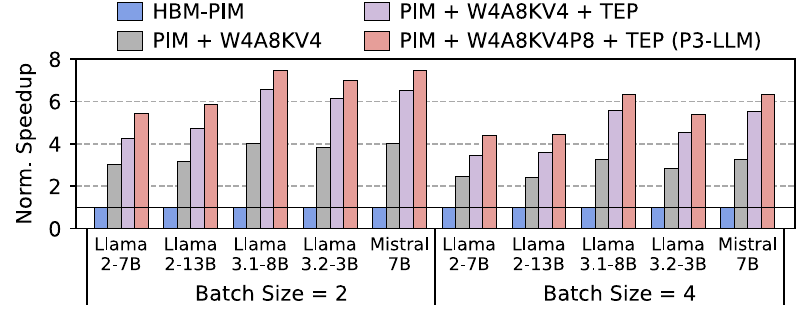}
        \vspace{-15pt}
        \caption{Ablation study on different architectural techniques of \workname{}. The context length is 4K.}
        \label{fig:arch_abl}
    \end{figure}
    

\vspace{3pt}
\noindent
\textbf{\labeltext{Architecture Ablation Study.}{pim_ablation}}
We conduct ablation studies to assess the performance gain of different architectural techniques proposed by \workname{}. Four designs are evaluated: (1) The baseline \text{HBM-PIM} accelerator running \texttt{FP16} models; (2) A PIM accelerator supporting \texttt{W4A8KV4} quantized models; (3) A PIM accelerator incorporating our throughput-enhanced PCU (TEP) to accelerate \texttt{W4A8KV4} models; (4) The proposed \workname{} with 8-bit attention-score on top of \texttt{W4A8KV4} and throughput-enhanced PCU. Fig.~\ref{fig:arch_abl} illustrates the normalized performance under batch sizes of 2 and 4. On average, \texttt{W4A8KV4} quantization achieves $3.3\times$ speedup over \text{HBM-PIM}, and adopting the throughput-enhanced PCU to exploit data reuse offers an additional $1.6\times$ speedup. 
\labeltext{Finally,}{p_quant_benefits} with 8-bit attention-score quantization, on average, \workname{} achieves another $1.2\times$ performance gain by enabling the low-precision PCU to fully accelerate the self-attention module.

    \begin{figure}[t]
        \centering
        \includegraphics[width=1\linewidth]{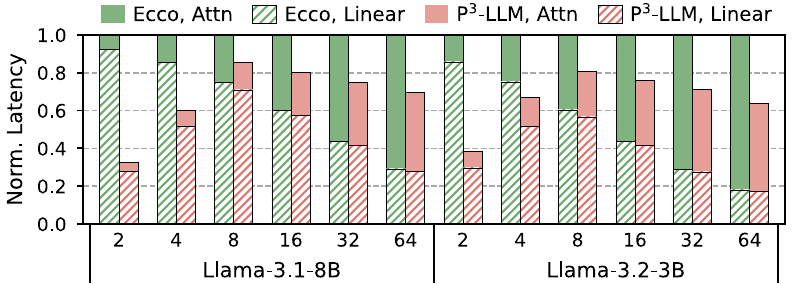}
        \vspace{-15pt}
        \caption{Normalized decoding latency ($\downarrow$) of Ecco and \workname{} across a wide range of batch sizes from 2 to 64. The context length is 4K.}
        \label{fig:large_batch}
    \end{figure}


\vspace{3pt}
\noindent
\textbf{\labeltext{Applicability to Large-Batch Decoding.}{reb:large_batch}} Although \workname{} mainly targets low-batch edge inference, it is also applicable to large-batch serving. To demonstrate this applicability, Fig.~\ref{fig:large_batch} presents the decoding latency breakdown of Ecco and \workname{} for two Llama-3 models across a wide range of batch sizes from 2 to 64. Notably, Ecco and \workname{} have similar latency for linear layers when the batch size reaches 8, as the PIM hardware becomes compute-bound. To address this, \workname{} offloads linear layers to NPU for more efficient execution. Interestingly, as the batch size continues to grow, \workname{} regains its performance advantage, which is attributed to the increasing dominance of attention layers in the overall runtime. Specifically, the GQA of Llama-3.1-8B and Llama-3.2-3B has a group size of 4 and 3, respectively, which still exhibits low data reuse opportunities. Thus, \workname{} can leverage its high internal PIM bandwidth and throughput-enhanced PCU to accelerate the fully quantized attention module, resulting in better performance than Ecco.

\section{Discussion} 

\noindent
\textbf{\labeltext{Broader Applicability of \workname{}.}{reb:discussion_ddr}} 
\textcolor{black}{
Apart from HBM-PIM, the proposed low-precision PCU of \workname{} is generally applicable to other DRAM variants such as GDDR~\cite{gddr-jedec} and LPDDR~\cite{lpddr-jedec}. For instance, the LPDDR-PIM~\cite{samsung-aim} introduced by Samsung adopts a similar PCU design as HBM-PIM, both containing a 16-way SIMD \texttt{FP16} MAC unit that receives 256-bit inputs and weights to compute GEMV. Furthermore, the throughput-enhanced PCU in \workname{} can also be applied to LPDDR, whose $t_{\text{CCD\_S}}$ is a half of $t_{\text{CCD\_L}}$.
}

\section{Related Works} 
\noindent
\textbf{\labeltext{Integer-based LLM Quantization.}{reb:related_work_int}} \textcolor{black}{
Numerous studies have explored quantization algorithms based on conventional integer formats to reduce the memory footprint and computational demand of LLMs~\cite{gptq, qserve, quarot, gyrot, tender, awq, smoothquant, llm-int8, kvquant, kivi, oaken, omniquant, spinquant, atom, bloom, flexquant, figna, lut-tc, earth, aster}. While weights are easy to quantize using integer formats, activations usually exhibit much larger dynamic range with significant quantization error. To address this, prior works employ various techniques, such as maintaining outliers at higher precision~\cite{oaken, llm-int8, atom}, smoothing activation channels~\cite{smoothquant, qserve}, and applying matrix rotation to activations~\cite{gyrot, quarot, spinquant}. However, these methods require either offline calibration with potential overfitting, or considerable online computation overhead. Our work distinguishes itself by seamlessly blending lightweight dynamic smoothing and hybrid numerical formats to achieve a balanced trade-off between model accuracy and hardware efficiency.
}

\vspace{3pt}
\noindent
\textbf{\labeltext{Custom Quantization Formats.}{reb:related_work_dtype}} \textcolor{black}{
Recently, there has been significant research on custom numerical formats for quantization in the architecture community~\cite{mant, bitmod, blockdialect, ant, microscaling, olive, mx_plus, spark, amove, microscopiq, anda, pisa, kllm, ecco, lut-gemm, figlut, razer, mxfp4_unveil, students}. Although these proposals achieve good hardware efficiency, they struggle to outperform existing algorithmic solutions under the aggressive \texttt{W4A8KV4} configuration~\cite{qserve, quarot}. On the other hand, \workname{} introduces a novel operand-dependent quantization scheme using hybrid numerical formats, achieving better accuracy than SoTA algorithmic solutions. 
In addition to academic proposals, custom quantization formats have been widely adopted by industry. The microscaling format~\cite{microscaling}, standardized by Open Compute Project, applies a shared 8-bit exponent to a group of low-precision operands. Both NVIDIA and AMD add support for \texttt{FP4} and \texttt{FP8} formats in their flagship GPUs~\cite{amd-gpu, blackwell-gpu}. The latest NVIDIA Blackwell architecture introduces the NVFP4 format~\cite{nvfp4}, which enhances the standard \texttt{FP4} data type with a two-level scaling strategy.
}

\vspace{3pt}
\noindent
\textbf{\labeltext{PIM-based LLM Inference Systems.}{reb:related_work_pim}} \textcolor{black}{
There is an abundance of works on LLM inference using NPU-PIM heterogeneous accelerators~\cite{neupim, pimba, ianus, duplex, plain, lp-spec, samsung-aim, sk-aim, attacc, orches, pimphony, aqpim, papi}. Samsung and SK Hynix have commercialized PIM products based on DDR and HBM technology~\cite{samsung-aim, sk-aim}. AttAcc~\cite{attacc}, IANUS~\cite{ianus}, and NeuPIMs~\cite{neupim} explore operator mapping and scheduling to improve the utilization of NPU and PIM during inference. These works focus exclusively on \texttt{FP16} LLM inference, making them orthogonal to our quantization method. 
Pimba~\cite{pimba}, PLAIN~\cite{plain}, and LP-Spec~\cite{lp-spec} design low-precision arithmetic units to enhance the throughput of PIM. However, they mainly target conservative quantization configurations such as \texttt{W8A8} and \texttt{W6A6}. In contrast, our work explores more aggressive \texttt{W4A8KV4P8} quantization to achieve better trade-off between model accuracy, memory footprint, and computational efficiency.
}

\section{Conclusion} 
In this paper, we present \workname{}, an algorithm-hardware co-design solution for efficient edge LLM inference on heterogeneous NPU-PIM accelerators. 
On the algorithm side, \workname{} develops a novel quantization scheme using hybrid numerical formats to quantize all major LLM operands, while maintaining superior model accuracy. 
On the hardware side, \workname{} is equipped with optimized low-precision dataflow and an area-efficient PCU architecture to support the proposed quantization scheme.
Our evaluation demonstrates that the \workname{} algorithm outperforms SoTA KV-cache-only and weight-activation quantization methods: Oaken, QuaRot, and QoQ.
Compared to SoTA LLM accelerators: HBM-PIM, Ecco, and Pimba, the \workname{} accelerator yields $4.9\times$, $2.0\times$, and $3.4\times$ speedups, respectively.

\section{Acknowledgment} 
We thank the anonymous reviewers of HPCA-2025 and ISCA-2026 for their helpful and constructive feedback. This work was supported in part by the National Science Foundation under Grant No. 2339084, the European Research Council (ERC) under grant agreement No. 101088865, the Flanders AI Research Program, and long-term structural Methusalem funding by the Flemish Government.



\clearpage
\vspace{5pt}
\bibliographystyle{IEEEtranS}
\bibliography{refs}

\end{document}